\documentclass[prx,twocolumn,superscriptaddress,amsmath,amsfonts]{revtex4}

\usepackage{graphicx,float}
\usepackage{braket}
\usepackage{siunitx}
\usepackage{amssymb}
\usepackage{mathrsfs,mathrsfs}
\usepackage{appendix}
\usepackage{xcolor}

\DeclareMathOperator{\sech}{sech}
\DeclareMathOperator{\Span}{span}
\begin{document}

\title{Broadband Parametric Downconversion as a Discrete-Continuum Fano Interaction}

\author{Ryotatsu Yanagimoto}
\thanks{These authors contributed equally to this work. \\Email: ryotatsu@stanford.edu, \\{\color{white}Email: }edwin98@stanford.edu}
\affiliation{E.\,L.\,Ginzton Laboratory, Stanford University, Stanford, California 94305, USA}

\author{Edwin Ng}
\thanks{These authors contributed equally to this work. \\Email: ryotatsu@stanford.edu, \\{\color{white}Email: }edwin98@stanford.edu}
\affiliation{E.\,L.\,Ginzton Laboratory, Stanford University, Stanford, California 94305, USA}

\author{Marc P. Jankowski}
\affiliation{E.\,L.\,Ginzton Laboratory, Stanford University, Stanford, California 94305, USA}
\affiliation{NTT Physics and Informatics Laboratories, NTT Research, Inc., 1950 University Ave., East Palo Alto, California 94303, USA}

\author{Tatsuhiro Onodera}
\affiliation{NTT Physics and Informatics Laboratories, NTT Research, Inc., 1950 University Ave., East Palo Alto, California 94303, USA}
\affiliation{School of Applied and Engineering Physics, Cornell University, Ithaca, New York 14853, USA}

\author{Martin M. Fejer}
\affiliation{E.\,L.\,Ginzton Laboratory, Stanford University, Stanford, California 94305, USA}

\author{Hideo Mabuchi}
\affiliation{E.\,L.\,Ginzton Laboratory, Stanford University, Stanford, California 94305, USA}

\date{\today}

\begin{abstract}
We introduce a theoretical framework based on Fano's theory of discrete-continuum interactions to analyze the quantum dynamics of broadband parametric downconversion (PDC) in the few-pump-photon regime of nonlinear quantum nanophotonics. Applying this unified analytic approach to 1D $\chi^{(2)}$-nonlinear waveguides, we find a host of remarkable dynamical features due to the coupling of a discrete pump state to the signal continuum, from unit-efficiency (i.e., complete) downconversion when the coupling is dissipative, to Rabi-like oscillations with sub-exponential decay when it is dispersive. The theory provides a straightforward way to analytically compute a full characterization of the PDC dynamics, including the complete eigensystem of the continuum Hamiltonian and expressions for the signal biphoton correlation function. We also apply the theory to study a pair of linearly coupled $\chi^{(2)}$ waveguides, where two discrete pump states simultaneously downconvert into a common-mode signal continuum, resulting in Fano interference that critically affects the PDC rate. Under appropriate conditions, the theory predicts characteristic Fano lineshapes and even complete destructive interference resulting in the full suppression of PDC, due to the formation of a bound pump state in the continuum. Generalizing further, we show that the framework can also be applied to higher-order parametric processes such as parametric three-photon generation, and we also find numerical signatures that Fano-type interactions occur even for multi-photon PDC under stronger pumping. Our results establish broadband PDC as yet another physical system natively exhibiting Fano-type interactions and advance a theoretical framework in which to understand the complicated quantum dynamics of strongly nonlinear broadband quantum optics.
\end{abstract}

\maketitle

\section{Introduction}
Harnessing the quantum nature of light potentially holds the key to overcoming classical limitations of conventional photonics in applications ranging from fundamental science, where coherent light sources have long been ubiquitous, to the more recent but rapidly developing field of quantum engineering and information processing~\cite{LIGO2013, Tsang2016, Kruse2016, Obrien2009}. In this context, quantum photonics stands out among other quantum-enhanced hardware platforms in its potential for long-distance connectivity, wide-bandwidth capacity, and room-temperature operability~\cite{Takeda2019, Rudolph2017, Asavarant2019, Yanagimoto2020}. Many state-of-the-art proposals for photonic quantum information processing are particularly reliant on the physical process of parametric downconversion (PDC)~\cite{Burnham1970, Loudon2000, Conteau2018}, in which a medium with some optical nonlinearity (e.g., $\chi^{(2)}$) facilitates the spontaneous generation of entangled photons. These entangled photons act as a key resource for a variety of quantum operations, and the non-classical correlations they carry are routinely exploited in foundational quantum photonic technologies like heralded single-photon generation~\cite{Motes2013,Pittman2005,Mosley2008}, quantum teleportation~\cite{Bouwmeester1997}, quantum key distribution~\cite{Adachi2007}, and precision measurement~\cite{Cheung2007,Cable2010}.

Nonlinear waveguides provide a convenient hardware platform for realizing the optical processes needed for PDC, as transverse spatial confinement allows better control over geometric dispersion and longer interaction lengths~\cite{Banaszek2001, Tanzilli2002, Eckstein2011}. However, in contrast to traditional quantum optics dealing with only a limited number of optical modes, as in single-mode cavity electrodynamics~\cite{Fink2008,Birnbaum2005}, waveguides are natively broadband systems supporting a large number of frequency modes, and a complete quantum model for PDC---fully incorporating the multimode and non-Gaussian aspects of its quantum dynamics---is computationally intractable in general~\cite{Lloyd1999, Niu2018, Braunstein2005}. Most established treatments of broadband PDC work in the weak-interaction limit where the PDC process is pumped by a strongly displaced coherent state, and crucially, remains undepleted (i.e., a negligible fraction of the pump energy is transferred in the process). This undepleted pump approximation produces a linearized, tractable quantum description of PDC appropriate for many experimental setups where the nonlinear interaction strength is small, and it has been useful for studying the effects of pulsed interactions~\cite{Grice1997}, spectral entanglement~\cite{Wasilewski2006}, waveguide loss~\cite{Helt2015}, and multi-photon emissions~\cite{Takeoka2015, Yang2008, Christ2013}.

On the other hand, significant recent progress in the fabrication of ultra-low-loss and highly nonlinear nanophotonic waveguides with precise phase and dispersion engineering~\cite{Wang2018, Zhang2017, Jankowski2020, Stanton2020,Chang2019, Lu2020} has significantly bridged the gap towards a ``strong-coupling regime'' for nonlinear photonics, where considerable energy transfer can occur between the pump and signal modes at the few-photon level, opening up the possibility for coherent generation of multimode non-Gaussian states in an all-photonic platform. Such a possibility underscores the importance of furthering our physical understanding, at both fundamental and engineering levels, of broadband quantum dynamics in this highly nonlinear regime, and of developing analytical and numerical paradigms for modeling potentially novel phenomena in such systems~\cite{Leung2009, Chang2008, Hafezi2012}.

In this research, starting from a full quantum model of a 1D $\chi^{(2)}$ waveguide, we show the emergence of exotic broadband quantum dynamics in the highly non-Gaussian regime beyond the undepleted pump approximation, such as perfect depletion of a weak pump field (i.e., unit-efficiency PDC) and Rabi-like oscillations with sub-exponential decay. We can uniformly analyze these broadband phenomena using an unconventional theoretical framework for PDC: Fano's theory for discrete-continuum interactions~\cite{Fano1961}, with which single-photon PDC is seen as an interaction between a discrete pump state and a continuum of signal states. In this picture, unit-efficiency PDC can be intuitively understood as a ``dissipation'' of a pump photon to a continuum composed of signal states via a process analogous to atomic/molecular autoionization~\cite{Madden1963, Wang2010, Geissler2001}; such phase decoherence within a closed system with many degrees of freedom is also reminiscent of intramolecular vibrational energy redistribution~\cite{Perez2018,Felker1985}. The qualitative nature of the Fano-type interaction for PDC can range from dissipative to dispersive depending on the system parameters, but by virtue of Fano's theory, analytical expressions for the system's dynamical behavior, including the time evolution of the pump photon population and biphoton signal correlation functions, can be derived for the entire parameter space in a unified way.

To illustrate how this unconventional approach results in both new intuition and powerful analytic tools for understanding potentially complicated nonlinear optical systems, we consider a simple example of PDC in two linearly coupled $\chi^{(2)}$ waveguides. By varying waveguide parameters and the initial pump state prepared as a superposition between the two waveguides, the discrete-continuum interactions in this system can destructively (constructively) interfere, causing a dramatic suppression (enhancement) of the overall PDC rate. As a result, this example already suffices to exhibit many of the rich physical phenomena that distinguishes Fano-type systems, including asymmetric Fano lineshapes in the continuum excitation spectrum and the formation of a bound state in the continuum (BIC). Analogously to many other studies of BIC~\cite{Hsu2016,Gomis-Bresco2017,Crespi2015,Longhi2007,Kodiagala2017,Minkov2018, Gorkunov2020}, this singular phenomenon can be seen as the existence of a protected pump state experiencing complete suppression of PDC due to destructive interference in its coupling to the signal-mode continuum.

In fact, discrete-continuum interactions likely underpin much of the physics of broadband parametric interactions in the few-pump-photon regime. For example, an analysis of three-photon generation proceeds analogously to the $\chi^{(2)}$ case, and the resulting phenomenology is qualitatively similar by virtue of the common applicability of Fano's theory. In addition, numerical simulations of $\chi^{(2)}$ PDC beyond the weak-excitation regime, taking into account multi-photon interactions, also show signatures of discrete-continuum interactions under appropriate conditions. These results suggest that the powerful toolbox of Fano's theory~\cite{Kroner2008, Limonov2017, Rybin2013, Harris1989, Tang2010, Miroshnichenko2010, Baksic2017} has much to offer for studying nonlinear broadband quantum optics. Such perspectives may provide useful intuition for qualitatively understanding complicated quantum dynamics even when numerical simulations are infeasible, in a spirit similar to the use of thermodynamic theory for highly multimode nonlinear optical systems~\cite{Wu2019}. Underscoring the importance of such theoretical tools, realistic experimental values for state-of-the-art nonlinear waveguide technology indicate the field of photonics is progressing towards this largely unexplored frontier.

\section{Quantum model of a broadband $\chi^{(2)}$ waveguide} \label{sec:model}
\begin{figure}[bh]
    \includegraphics[width=0.45\textwidth]{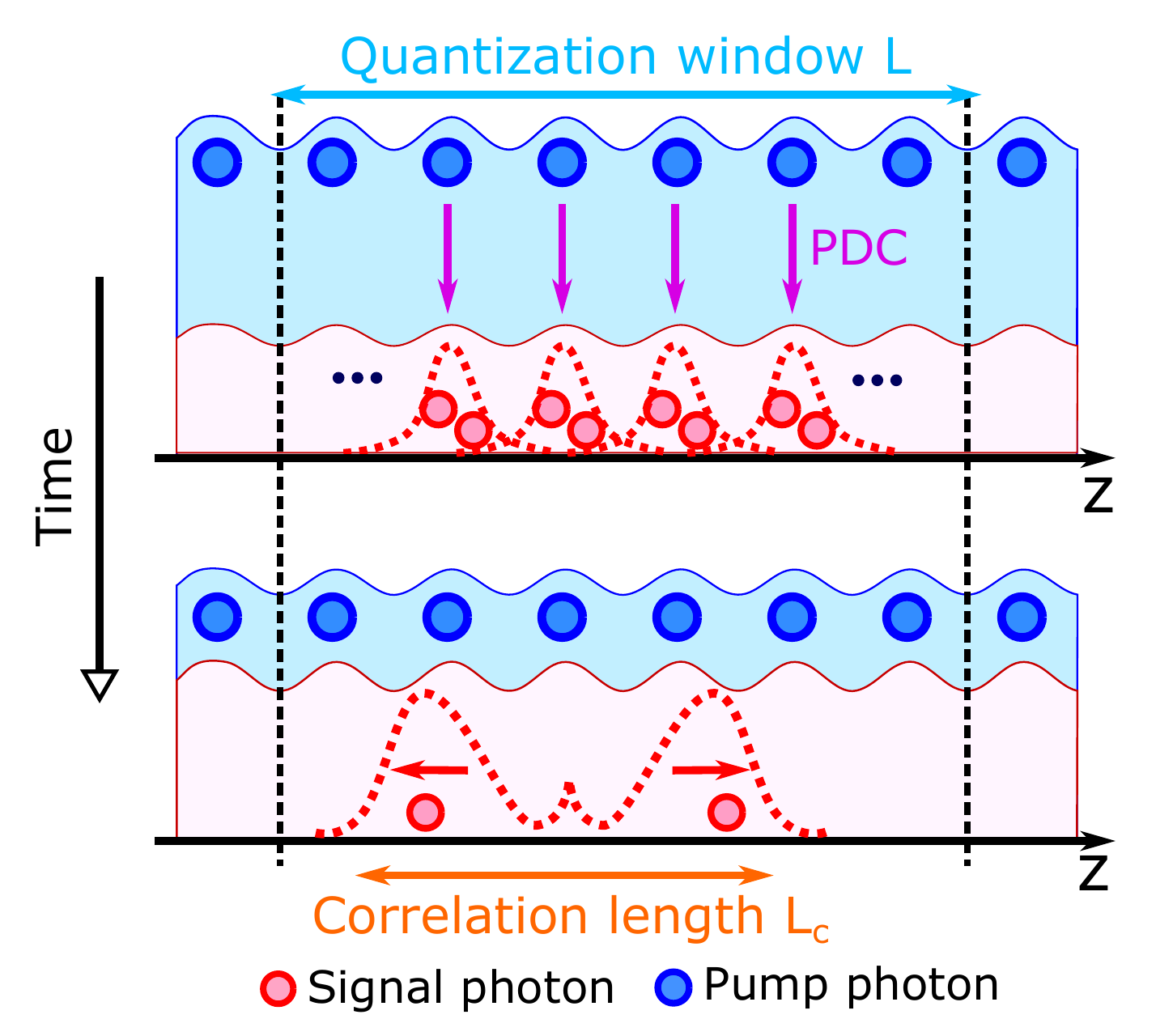}
    \caption {Schematic representation of the length scales involved in our quantization procedure, for the case of parametric downconversion of a monochromatic pump field. Pump photons (blue circles) downconvert via a $\chi^{(2)}$ interaction, which generates localized signal photon pairs (red circles) as shown in the upper figure. Red dashed lines schematically indicate the spatial extent of the signal biphoton correlation functions. After some time (lower figure), the signal photons undergo dispersion, resulting in some spatial correlation structure with a length scale $L_\text{c}$. To quantize the system, we apply periodic boundary conditions to a quantization window of length $L$, chosen large enough to contain $L_\text{c}$.}
    \label{fig:quantization}
\end{figure}

In this section, we construct a quantum model for a dispersive 1D $\chi^{(2)}$ nonlinear waveguide based on the canonical quantization procedure developed in Ref.~\cite{Drummond1990, Drummond2014}. We consider a quantization window of length $L$ to which periodic boundary conditions are applied, as shown in Fig.~\ref{fig:quantization}, and the length $L$ should be chosen long enough to avoid boundary effects on the dynamics. In PDC, for instance, the spatial distribution and correlation function of the downconverted signal photon pairs have length scales which undergo dispersion as a function of time. If the maximum such length scale of interest is $L_\text{c}$, then we must take $L>L_\text{c}$. We discuss $L_\text{c}$ (in terms of the biphoton correlation function, etc.) more quantitatively in Sec.~\ref{sec:spatialcorrelation}.

To treat PDC, we assume the field can be partitioned into a signal (i.e., fundamental) band of frequencies centered around some wavevector $k_{\text{a}0}$ and a pump (i.e., second-harmonic) band around $k_{\text{b}0} = 2k_{\text{a}0}$. Under periodic boundary conditions, the quantization window supports discrete wavevectors $k_{\text{a}m} = k_{\text{a}0} + (2\pi/L)m$ and $k_{\text{b}\ell} = k_{\text{b}0} + (2\pi/L)\ell$ ($m,\ell \in \mathbb Z$) for the signal and pump bands, respectively. We also assume that broadband excitations of the waveguide have bandwidths which are primarily limited by second-order dispersion, so that a monochromatic mode with wavevector $k$ and energy $\hbar \omega(k)$ has a dispersion relation of the form $\omega(k) = \omega(k_0) + \omega'(k_0) (k-k_0) + \frac12 \omega''(k_0) (k-k_0)^2$, with all higher-order dispersion neglected.

To quantize the field, we introduce photon-polariton field annhilation operators $\hat{a}_m$ and $\hat{b}_\ell$ for the signal and pump fields~\cite{Raymer2020,Quesada2017}, respectively, which are defined in rotating frames rotating at $\omega(k_{\text{a}0}) + (k_{\text{a}m}-k_{\text{a}0})\omega'(k_{\text{a}0})$ and $2\omega(k_{\text{a}0}) + (k_{\text{b}\ell}-2k_{\text{a}0})\omega'(k_{\text{a}0})$, respectively.  With this, the system is described by a rotating-frame Hamiltonian
\begin{align} \label{eq:originalh}
    \hat{H}/\hbar&=\frac{\eta^{(2)}A_\mathrm{eff}}{\hbar}\int_L \mathrm{d}z\,\bigl(\partial_z\hat{\Lambda}_\text{a}\bigr)^2\partial_z\hat{\Lambda}_\text{b} \\
    &{}+\sum^\infty_{m=-\infty}d_\text{a} m^2\hat{a}    _{m}^\dagger\hat{a}_{m}+\sum^\infty_{\ell=-\infty}(\delta+\mu \ell+d_\text{b}\ell^2)\hat{b}_{\ell}^\dagger\hat{b}_{\ell}. \nonumber
\end{align}
In the nonlinear part, $\eta^{(2)}=-\chi^{(2)}/\epsilon_0^2n_\text{a}^4n_\text{b}^2$ where $\chi^{(2)}$ is the effective second-order permittivity~\cite{Quesada2017}, $n_\text{a}$ and $n_\text{b}$ are the indices of refraction for the signal and pump, and $A_\text{eff}$ is the effective interaction area of the modes~\cite{Jankowski2020}; all these quantities are assumed to be frequency-independent. In the dispersive part, $\delta$ is the relative detuning between pump and signal carriers (i.e., the phase mismatch), $\mu$ is the linear dispersion of the pump due to group velocity mismatch (in the comoving frame of the signal carrier), and $d_a$ and $d_b$ are the quadratic dispersions of the signal and pump carriers due to group velocity dispersion; these constants are given by
\begin{subequations}
\label{eq:dispersions}
\begin{align}
    \delta&=\omega(k_{\text{b}0})-2\omega (k_{\text{a}0}) \\
    \mu&=\frac{2\pi}{L}\bigl(\omega'(k_{\text{b}0})-\omega' (k_{\text{a}0})\bigr) \\
    d_\text{a}&=\left(\frac{2\pi}{L}\right)^2\frac{\omega''(k_{\text{a}0})}{2};\quad d_\text{b}=\left(\frac{2\pi}{L}\right)^2\frac{\omega''(k_{\text{b}0})}{2}.
\end{align}
\end{subequations}
The dual potentials $\hat\Lambda_\text{a}$ and $\hat\Lambda_\text{b}$, which are related to the displacement field operators via $\hat{\boldsymbol{D}}=\nabla \times\hat{\boldsymbol{\Lambda}}$, constitute the canonical coordinates of our quantization procedure and are given in the rotating frame by
\begin{subequations}
\begin{align}
    &\hat{\Lambda}_\text{a}=\sum^\infty_{m=-\infty}\sqrt{\frac{\hbar \epsilon_0n_\text{a}\omega'(k_{\text{a}m})}{2k_{\text{a}m}A_\text{a}L}} \, \hat{a}_{m} e^{\mathrm{i}k_{\text{a}m}z} \label{eq:lambdaA} \\
    &\quad\times \exp\Bigl(-\mathrm{i}\Bigl[\omega(k_{\text{a}0})+(k_{\text{a}m}-k_{\text{a}0})\omega'(k_{\text{a}0})\Bigr]t\Bigr)+\text{H.c.} \nonumber \\
    &\hat{\Lambda}_\text{b}=-\mathrm{i}\sum^\infty_{\ell=-\infty}\sqrt{\frac{\hbar \epsilon_0n_\text{b}\omega'(k_{\text{b}\ell})}{2k_{\text{b}\ell}A_\text{b}L}} \, \hat b_\ell e^{\mathrm{i}k_{\text{b}\ell}z} \label{eq:lambdaB} \\
    &\quad\times \exp\Bigl(-\mathrm{i}\Bigl[2\omega(k_{\text{a}0})+(k_{\text{b}\ell}-2k_{\text{a}0})\omega'(k_{\text{a}0})\Bigr]t\Bigr)+\text{H.c.}, \nonumber
\end{align}
\end{subequations}
where $A_\text{a}$ and $A_\text{b}$ are the transverse cross-section areas of the signal mode and the pump modes, respectively (which are also assumed to be independent of frequency).

We now insert \eqref{eq:lambdaA} and \eqref{eq:lambdaB} into \eqref{eq:originalh} and perform the integral over $z$. After also applying a rotating wave approximation to eliminate terms rotating at frequencies on the order of $\omega(k_\text{a,0})$, we obtain
\begin{align} \label{eq:fullhamiltonian}
    \hat{H}/\hbar &= \frac{g}{2}\sum_{m+n=\ell}\left(\hat{a}_{m}^\dagger\hat{a}_{n}^\dagger\hat{b}_{\ell}+\hat{a}_{m}\hat{a}_{n}\hat{b}_{\ell}^\dagger\right) \\
    &{}+\sum^\infty_{m=-\infty}d_\text{a} m^2\hat{a}
_{m}^\dagger\hat{a}_{m} + \sum_{\ell=-\infty}^\infty (\delta+\mu \ell+d_\text{b}\ell^2)\hat{b}_{\ell}^\dagger\hat{b}_{\ell}, \nonumber
\end{align}
where the nonlinear coupling rate is
\begin{align}
    g = \chi^{(2)}\sqrt{\frac{\hbar \omega'{}^2(k_{\text{a}0})\omega'{}
    (k_{\text{b}0})k_{\text{a}0}^3A_\text{eff}^2 }{\epsilon_0n_\text{a}^4n_\text{b}^2A_\text{a}^2A_\text{b}L}}.
\end{align}

Note that the Hamiltonian \eqref{eq:fullhamiltonian} commutes with both the generalized number operator $\hat{N}=\frac{1}{2}\hat{N}_\text{a}+\hat{N}_\text{b}$, where $\hat{N}_\text{a}=\sum_{m=-\infty}^\infty \hat{a}_m^\dagger\hat{a}_m$ and $\hat{N}_\text{b}=\sum_{\ell=-\infty}^\infty \hat{b}^\dagger_\ell\hat{b}_\ell$, as well as the total momentum operator $\hat{M}=\sum_{m=-\infty}^\infty m\hat{a}_m^\dagger\hat{a}_m+\sum_{\ell=-\infty}^\infty\ell\hat{b}_\ell^\dagger\hat{b}_\ell$; i.e., $[\hat{H},\hat{N}]=[\hat{H},\hat{M}]=0$. Thus, $\hat N$ and $\hat M$ represent conserved quantities of the evolution under \eqref{eq:fullhamiltonian}, and we can decompose the entire system Hilbert space into non-interacting eigenspaces of $\hat{N}$ and $\hat{M}$.

\section{PDC in the weak-excitation regime} \label{sec:singlephotonPDC}
Although the general quantum dynamics under the Hamiltonian \eqref{eq:fullhamiltonian} are intractable due to the immense size of the Hilbert space, the special case of single-photon PDC offers a good deal of analytic and qualitative insight into the behavior of the system. Notably, the results obtained for single-photon PDC describe not only the case of a single input pump photon but also the dynamics experienced by a weak coherent pump field. To see this, consider a monochromatic coherent state with wavevector $k_{b\ell}$ and photon density $\rho$ (related to the photon flux in a non-comoving frame). When $\rho\ll 1/L_\text{c}$, there is, intuitively, fewer than one photon per correlation length. We refer to this as the weak-excitation regime, since this condition allows us to take the length $L$ of the quantization window to satisfy $\rho L \ll 1$ while still satisfying $L_\text{c} < L$. In this case, the initial state of the system is $\exp\bigl[\sqrt{\rho L}(\hat{b}_\ell^\dagger-\hat{b}_\ell)\bigr]\ket{0} \propto \ket{0}+\sqrt{\rho L}\,\hat{b}_\ell^\dagger \, \ket{0} + \mathcal{O}(\rho L)$, where the contribution from the higher-order terms can be neglected. Since the vacuum component evolves trivially, the evolution of the single-photon component suffices to fully capture the dynamics of the input coherent state.

The state $\hat{b}_\ell^\dagger\ket{0}$ represents the existence of a single pump photon somewhere in the quantization length $L$. Since such a situation cannot exhibit multi-photon interactions in the pump, the dynamics of single-photon PDC is expected to be independent of $L$ (i.e., independent of the pump photon ``density''). (As discussed in Sec.~\ref{sec:general}, this is not the case for coherent-state inputs outside the weak-excitation regime as we then have to consider states such as $\hat b_\ell^{\dagger2}\ket{0}$, whose dynamics are dependent on $L$.) As we will show, this invariance of single-photon PDC to rescalings in $L$ allows us to take a continuum limit $L\rightarrow\infty$, allowing us to approximate the energy spectrum as continuous and providing useful analytic results to describe PDC in the weak-excitation regime.

In the continuum limit, the system Hamiltonian exhibits Fano-type discrete-continuum interactions. Thus, using Fano theory, we can obtain complete eigenspectra along with the corresponding eigenstates, leading to analytical expressions for the system dynamics, including the time evolution of the pump photon population and the spatial biphoton correlation function of the downconverted signal photons. It is shown that the nature of the discrete-continuum interaction can vary from dissipative to dispersive, depending on the detuning of the pump mode. In the dissipative regime, near-perfect PDC is realized via an analogous process to atomic/molecular autoionization. In the dispersive regime, the pump photon population exhibits Rabi-like oscillations with sub-exponential decay. We compare these analytic results derived with the continuum model (with $L\rightarrow\infty$) against numerical results from the original discrete Hamiltonian with finite $L$ to validate the approach.

\subsection{Single-photon PDC}
\label{sec:discrete}
Let us consider the initial state to be a single-photon excitation of a monochromatic pump field with wavevector $k_{\text{b}\ell}$,
\begin{align}
    \ket{b_\ell} = \hat{b}_{\ell}^\dagger\ket{0}.
\end{align}
Starting from this initial state, the evolution of the system is in fact closed within a subspace spanned by $\ket{b_\ell}$ and a band of signal states (consisting of downconverted pairs) given by
\begin{align}
    \ket{a_{p,\ell}} = \begin{cases}
        \frac{1}{\sqrt{2}}\hat{a}_0^\dagger{}^2\ket{0} & p=\ell=0 \\
        \hat{a}^\dagger_{\lceil \ell/2 \rceil+p}\hat{a}^\dagger_{\lfloor \ell/2 \rfloor-p}\ket{0} & \text{otherwise}
    \end{cases},
\end{align}
where the index $p \in \{0,1,\ldots\}$ corresponds to the ``non-degeneracy'' of the downconverted signal state relative to the input pump state. Within this subspace, the nonzero matrix elements are
\begin{subequations} \begin{align} \label{eq:hlelements}
    &\braket{b_\ell|\hat{H}|b_\ell}/\hbar = \delta + \mu \ell + d_b\ell^2 \\
    \label{eq:signaldiag}
    &\braket{a_{p,\ell}|\hat{H}|a_{p,\ell}}/\hbar =2d_\text{a}\begin{cases} \left(\textstyle\frac\ell2\right)^2+\left(p\right)^2 & \ell \in \text{even}\\
    \left(\textstyle\frac\ell2\right)^2+\left(p+\textstyle\frac12\right)^2 & \ell \in \text{odd}
    \end{cases}\\
    \label{eq:offdiag}
    &\braket{a_{p,\ell}|\hat{H}|b_{\ell}}/\hbar = \begin{cases}
    \frac{1}{\sqrt{2}}g & p = \ell = 0 \\
    g & \text{otherwise}
    \end{cases}
\end{align} \end{subequations}

Investigating these equations, we first see that when $\ell$ is odd, the only effect is a shift of $p$ by $1/2$. Therefore, we may consider solutions for even $\ell$ and arbitrary $p$ without loss of generality. Second, for a given $\ell$, the net effect of $\mu$ and $d_\text{b}$ is to shift the detuning $\delta$, and thus, we may consider the case $\mu = d_\text{b} = 0$ without loss of generality. Furthermore, we also see that we can subtract off a constant offset $2d_\text{a}(\ell/2)^2$ from the diagonal elements in \eqref{eq:hlelements} and \eqref{eq:signaldiag}, upon which we find that the only dependence on the index $\ell$ comes from \eqref{eq:hlelements}. Thus, the nonzero momentum index $\ell$ in the initial pump state is, up to a constant energy offset, again equivalent to a shift of the detuning $\delta$, and we can focus on the case of $\ell = 0$ without loss of generality. Together this means the only non-trivial dynamics of the system occurs for $\mu = d_\text{b} = 0$ and arbitrary $\delta$, $d_\text{a}$, and $g$, and we can restrict the Hilbert space to the $\ell = 0$ subspace with arbitrary $p$, reachable from the ``dc'' (with respect to the pump carrier) initial state $\ket{b_0}$. Based on these results, propagation of a single-photon pump \emph{pulse} with general form $\sum_{\ell=-\infty}^\infty f_\ell \ket{b_\ell}$ can be obtained by simply solving the PDC dynamics of each monochromatic component $\ket{b_\ell}$.

Because the choice of the quantization window length $L$ is arbitrary, we should be able to find a timescale for these dynamics which is independent of $L$. Through dimensional analysis, it turns out that the rate
\begin{align}
\label{eq:kappa}
    \kappa = \left(\frac{g^4}{2d_\text{a}}\right)^\frac{1}{3}
\end{align}
satisfies this property, and it effectively plays the role of the effective nonlinear coupling rate in this dimensionalization. We also define a dimensionless parameter $\xi = \delta/\kappa$, which corresponds to the effective (normalized) detuning. Then together with the above simplifications, the matrix elements of interest are
\begin{subequations} \label{eq:dchamiltonian}
\begin{align}
    \label{eq:dcpumpdiag}
    \braket{b_{0}|\hat{H}|b_{0}}/\hbar\kappa &= \xi \\
    \label{eq:dcsignaldiag}
    \braket{a_{p,0}|\hat{H}|a_{p,0}}/\hbar\kappa &= \epsilon^2p^2 \\
    \label{eq:dcoffdiag}
    \braket{a_{p,0}|\hat{H}|b_{0}}/\hbar\kappa &= \epsilon^\frac{1}{2},
\end{align}
\end{subequations}
where the dimensionless parameter $\epsilon = (2d_\text{a}/g)^{\frac{2}{3}} \propto L^{-1}$; nonzero $\epsilon$ is a consequence of choosing a finite window length $L$, and in the continuum limit of $L \rightarrow \infty$, $\epsilon \rightarrow 0$. Note, however, this does not mean that the matrix elements in \eqref{eq:dcsignaldiag} and \eqref{eq:dcoffdiag} become negligible in this limit, since the number of such elements (i.e., the density of the band of signal states) is also increasing with $L$. In fact, as we show in the following, the dynamics under \eqref{eq:dchamiltonian} become independent of $L$ yet remain non-trivial in the limit $\epsilon\rightarrow 0$. In order to retain $\epsilon > 0$, we also assume in the following that $d_\text{a}>0$ for simplicity; the dynamics for $d_\text{a}<0$ can be recovered by considering $\delta\mapsto -\delta$, $t\mapsto -t$ (due to $\kappa \mapsto -\kappa$), and $\hat{b}_0 \mapsto -\hat{b}_0$ (and a similar procedure can be generally done on the level of \eqref{eq:fullhamiltonian} including the non-dc modes as well).

The Hamiltonian matrix elements \eqref{eq:dchamiltonian} can be seen as describing the hopping of a single excitation among the set of signal states together with the single dc pump state. As a result, we can summarize the dynamics by introducing two-level lowering operators
\begin{align}
\hat{v}=\ket{0}\bra{b_0},
\end{align}
which annihilates the dc pump excitation, and
\begin{align}
    \hat{u}_p &= \ket{0}\bra{a_{p,0}} 
\end{align}
which annihilates a photon-pair excitation with non-degeneracy $p$. Using these operators, we can define an effective hopping Hamiltonian for this subspace, normalized with respect to $\hbar\kappa$, as
\begin{align} \label{eq:ghamiltonian}
    \hat{G} = \xi \hat{v}^\dagger\hat{v}+\sum^\infty_{p=0}\left[\epsilon^{2}p^2\hat{u}_p^\dagger\hat{u}_p+\epsilon^\frac{1}{2}\left(\hat{u}_p^\dagger\hat{v}+\hat{u}_p\hat{v}^\dagger\right)\right],
\end{align}
which manifestly has the matrix elements in \eqref{eq:dchamiltonian}.

The form of the Hamiltonian \eqref{eq:ghamiltonian} is well-suited for numerical simulation of the dynamics of single-photon PDC. By picking a finite $L$ (large enough to eliminate boundary effects) and an appropriate momentum cutoff $p_\text{max}$ (such that $p < p_\text{max}$), we arrive at an efficient numerical approximation for the dynamics, with Hilbert space dimension scaling as $p_\text{max}$. (The required $p_\text{max} \sim L$ but also depends on $\xi$; we elaborate on the latter scaling in the next subsection.) More specifically, $\hat{G}$ can be approximated by a matrix
\begin{align}
\hat{G}\approx\begin{pmatrix}
\xi&\epsilon^\frac{1}{2}&\epsilon^\frac{1}{2}&\epsilon^\frac{1}{2}&\dots&\epsilon^\frac{1}{2}\\
\epsilon^\frac{1}{2}&0&0&0&\dots&0\\
\epsilon^\frac{1}{2}&0&\epsilon^2&0&\dots&0\\
\epsilon^\frac{1}{2}&0&0&2^2\epsilon^2&\dots&0\\
\vdots&\vdots&\vdots&\vdots&\ddots&
\vdots\\
\epsilon^\frac{1}{2}&0&0&0&\dots& p_\text{max}^2\epsilon^2
\end{pmatrix},
\end{align}
where the first dimension corresponds to $\ket{b_0}$, while $(p+2)$th dimension corresponds to $\ket{a_{p,0}}$.

\subsection{Eigenstates in the continuum limit}
\label{sec:continuum}
In order to derive additional analytic insight into single-photon PDC, we would like to make an analogy to the physics of Fano-type discrete-continuum interactions~\cite{Fano1961}. For this purpose, we first need to take the continuum limit of $L \rightarrow \infty$ and work with continuum fields. Consider a rescaled momentum $s=\epsilon p$ (essentially a non-dimensionalized wavevector difference between the two downconverted photons), together with the set of field operators
\begin{align}
    \hat{\phi}_s=\epsilon^{-\frac{1}{2}}\hat{u}_p.
\end{align}
In the limit of $L\rightarrow \infty$ (equivalently $\epsilon\rightarrow 0$), $s$ becomes a continuous coordinate, while the commutation relationships for $\hat{\phi}_s$ obeys $[\hat{\phi}_s,\hat{\phi}^\dagger_{s'}]=\delta(s-s')$. Using these substitutions, we can derive that $\lim_{L\rightarrow\infty} \hat G = \hat G^\text{cont}$, where
\begin{align}
    \label{eq:gcont}
    \hat{G}^\mathrm{cont} = \xi\hat{v}^\dagger\hat{v}+\int_0^\infty \mathrm{d}s \left(s^2\hat{\phi}_s^\dagger\hat{\phi}_s+\hat{v}\hat{\phi}_s^\dagger+\hat{v}^\dagger\hat{\phi}_s\right).
\end{align}
As expected, $\hat G^\text{cont}$ only depends on $\xi$, which is independent of $L$.

\begin{figure}[ht]
\centering
    \includegraphics[width=0.48\textwidth]{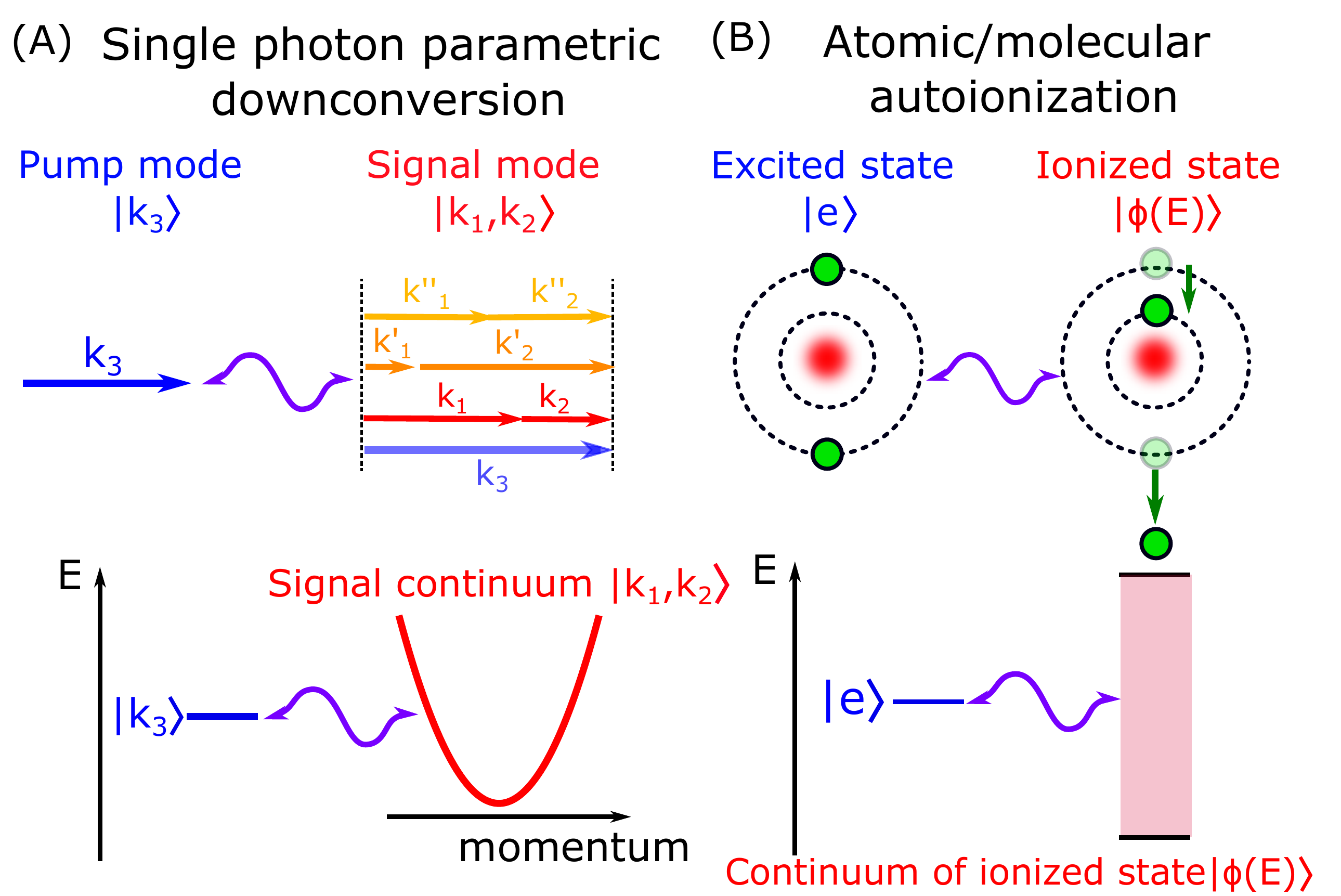}
    \caption{(A) Single-photon parametric downconversion happens so that the momentum is conserved. For each pump mode with momentum $k_3$, there exists a continuum of signal modes to couple to. (B) Autoionization of an atom. When the energy of the excited state $\ket{e}$ lies within the continuum of ionized state $\ket{\phi(E)}$, the atom ionizes unitarily.}
    \label{fig:autoionization}
\end{figure}

This Hamiltonian has a characteristic structure of Fano-type interaction in which a discrete pump state $\hat{v}^\dagger\ket{0}$ is coupled to a continuum of signal states $\hat{\phi}_s^\dagger\ket{0}$. As shown in Fig.~\ref{fig:autoionization}(A), when a monochromatic single-photon pump state $\ket{k_3}_\text{b}$ with wavevector $k_3$ downconverts to a pair of signal photons $\ket{k_1,k_2}_\text{a}$, it can do so with any continuous combination of wavevectors $k_1$ and $k_2$, so long as it satisfies momentum conservation $k_1+k_2=k_3$. Depending on the choice of the system detuning and dispersion, the energy of $\ket{k_3}_\text{b}$ can lie within or outside of the continuous energy band of the states $\ket{k_1,k_2}_\text{a}$, leading to dissipative or dispersive coupling, respectively. Energetically, this dissipative coupling to the continuum is reminiscent of atomic systems with configuration interactions that exhibit autoionization~\cite{Madden1963, Wang2010, Geissler2001}: When the energy of a discrete excited state of an atom (a molecule) lies within the continuous energy band of the ionized states, as shown in Fig.~\ref{fig:autoionization}(B), an atom (a molecule) initially in the excited state ionizes unitarily through a process known as autoionization. We show that a direct mapping can be established between \eqref{eq:gcont} and atomic systems, and phenomena such as the ``autoionization'' of pump photons occur as a direct consequence of this mapping. This autoionization of the pump photons provides an efficient channel for downconversion (i.e., with near-complete pump depletion).

Following Ref.~\cite{Fano1961}, we posit that the eigenstates of $\hat{G}^\mathrm{cont}$ with eigenvalue $\lambda$ take the form
\begin{align}
\label{eq:eigen}
    \ket{\varphi_\lambda}=\left(c_\lambda \hat{v}^\dagger+\int^\infty_0\mathrm{d}sf_\lambda(s)\hat{\phi}_s^\dagger\right)\ket{0}. 
\end{align}
By virtue of $\hat{G}^\text{cont}\ket{\varphi_\lambda}=\lambda\ket{\varphi_\lambda}$, $c_\lambda$ and $f_\lambda$ have to satisfy
\begin{subequations} \begin{align}
    \label{eq:clambda}
    \xi c_\lambda+\int^\infty_0\mathrm{d}sf_\lambda(s)&=\lambda c_\lambda\\
    \label{eq:flambda}
    c_\lambda+s^2f_\lambda(s)&=\lambda f_\lambda(s).
\end{align} \end{subequations}

We first consider the case of $\lambda < 0$, with $\lambda = -\lambda_\text{M}$ (the reason for the M-subscript will become clear later). In this case, the solution to \eqref{eq:flambda} takes the form
\begin{align}
    f_{-\lambda_\text{M}}(s) = -\frac{c_{-\lambda_\text{M}}}{\lambda_\text{M}+s^2},
\end{align}
and substituting this expression into \eqref{eq:clambda} leads to the condition
\begin{align} \label{eq:implicit}
    \frac{\pi}{2\sqrt{\lambda_\text{M}}} - \lambda_\text{M} = \xi,
\end{align}
which can be implicitly solved to determine $\lambda_\text{M}$ as a function of $\xi$. Importantly, there exists only one solution to \eqref{eq:implicit} satisfying $\lambda_\text{M} > 0$, which means that this is a unique, discrete state.  Thus, we define $f_{\text{M}}(s) = f_{-\lambda_\text{M}}(s)$, $c_{\text{M}} = c_{-\lambda_\text{M}}$ and $\ket{\varphi_\text{M}} = \ket{\varphi_{-\lambda_\text{M}}}$. Finally, imposing the normalization condition $\braket{\varphi_{\text{M}}|\varphi_{\text{M}}} = 1$ to this discrete state gives
\begin{align} \label{eq:cm}
c^2_{\text{M}} = \left(1+\frac{\pi}{4\lambda_\text{M}^{3/2}}\right)^{-1}.
\end{align}
Notably, this state turns out to be an ``optical meson'' state (hence the M-subscript), which was originally discovered in Ref.~\cite{Drummond1997}. Such photon bound states have been observed experimentally utilizing nonlinearities induced by atomic vapor~\cite{Liang2018,Firstenberg2013}. Based on \eqref{eq:implicit} and \eqref{eq:cm}, we find $\lim_{\xi\rightarrow -\infty}\mathrm{c}_{\text{M}}^2=1$ while $\lim_{\xi\rightarrow \infty}\mathrm{c}_{\text{M}}^2=0$. This means that for large negative values of the detuning $\xi$, the optical meson state is composed mostly of pump-photon excitation, while for large positive values of the detuning, the state is composed mostly of downconverted signal pairs.

Considering now the other case, when $\lambda \geq 0$, \eqref{eq:flambda} has a singularity at $s^2 = \lambda$. To handle this case, we write the solution to \eqref{eq:flambda} in the form
\begin{align}
    \label{eq:flambda_z}
    f_{\lambda}(s) = c_{\lambda}\left(\frac{1}{\lambda-s^2}+w(\lambda)\delta(\lambda-s^2)\right)
\end{align}
where $w(\lambda)$ is to be determined in order to satisfy \eqref{eq:clambda}. By substituting \eqref{eq:flambda_z} into \eqref{eq:clambda} and taking the Cauchy principal value of the integral, we find
\begin{align}
    w(\lambda)=2\sqrt{\lambda}(\lambda-\xi).
\end{align}
Finally, since we now have a continuous set of solutions $\lambda$, the normalization condition determining $c_\lambda$ is taken to be $\braket{\varphi_{\lambda}|\varphi_{\lambda'}}=\delta(\lambda-\lambda')$, which results in
\begin{align}
\label{eq:clambdacont}
   c_{\lambda}^2=\frac{2\sqrt{\lambda}}{w^2(\lambda)+\pi^2}.
\end{align}
Notice that for large enough $\xi>0$ (i.e. dissipative coupling), \eqref{eq:clambdacont} takes a Lorentzian lineshape centered around $\lambda=\xi$, reflecting that discrete pump state whose unperturbed energy being $\xi$ is strongly coupled to the continuum with similar energy.

\iffalse
\begin{align}
    \lambda_-=\frac{\left(\sqrt[3]{8 \xi ^3+3 \pi  \sqrt{48 \xi ^3+81 \pi ^2}+27 \pi ^2}-2 \xi \right)^2}{6 \sqrt[3]{8 \xi ^3+3 \pi  \sqrt{48 \xi ^3+81 \pi ^2}+27 \pi ^2}}.
\end{align}
\fi

\subsection{Dynamics of pump photon population}
\begin{figure}[hb]
\centering
    \includegraphics[width=0.43\textwidth]{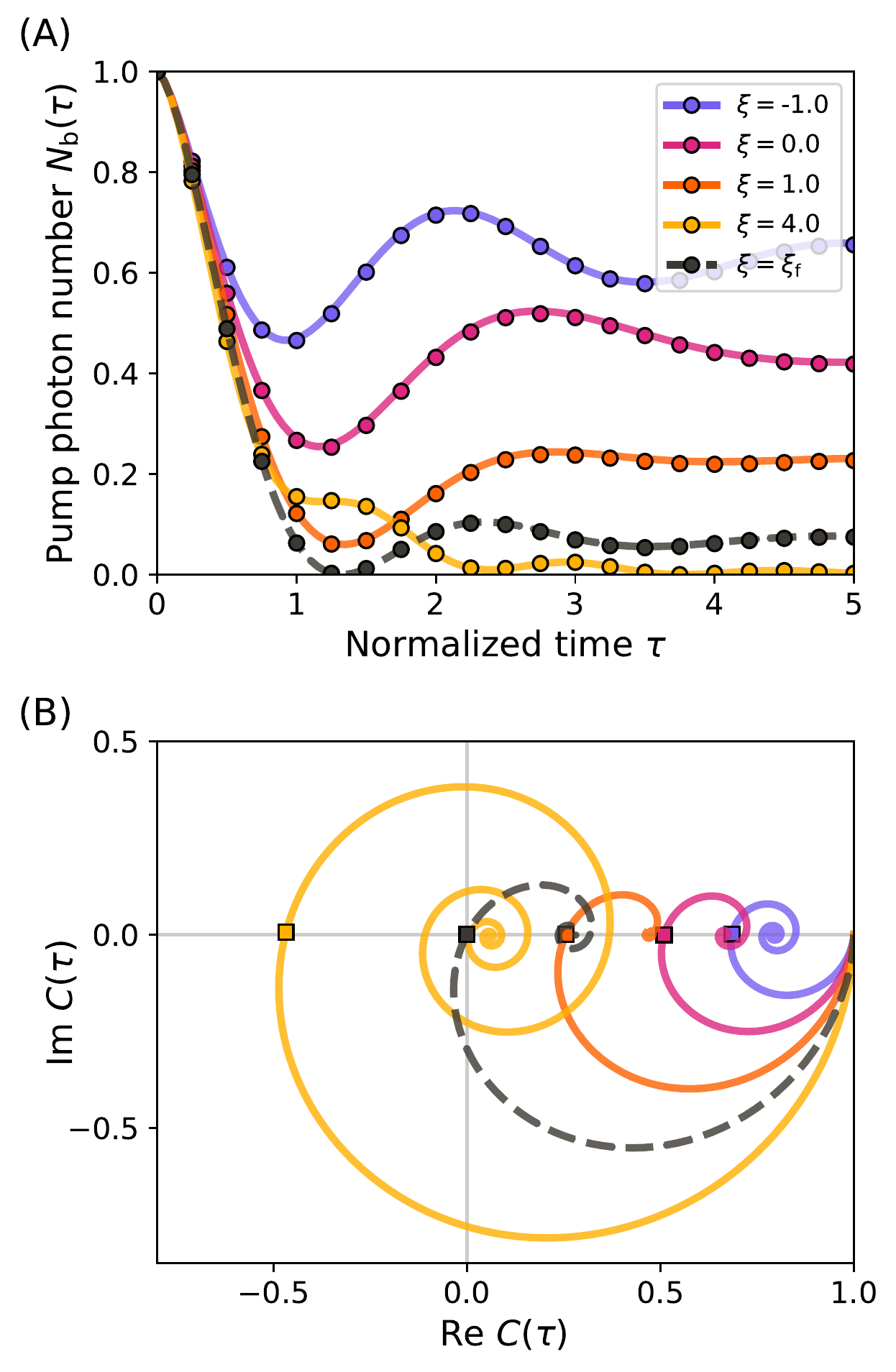}
    \caption{(A) Pump photon population $N_\text{b}(\tau)$ as a function of (normalized) propagation time $\tau$ for various (normalized) detunings $\xi$ (i.e., phase mismatch), ranging from dispersive ($\xi < 0$) to dissipative ($\xi > 0$). Solid lines are based on analytic evaluation of \eqref{eq:evolution}, while circles show numerical results based on simulating \eqref{eq:ghamiltonian} with finite quantization window length ($\epsilon=1/30$). (B) Trajectories taken by the pump amplitude $C(\tau)$ ($|C|^2 = N_\text{b}$) in the complex plane for various $\xi$. Squares represent the point where $C(\tau)$ crosses the real axis for the first time after $\tau=0$. As discussed in the main text, finite-time pump depletion can also happen; here, $N_\text{b} = C = 0$ for $\xi = \xi_\text{f}\approx 1.90$ at $\tau = \tau_\text{f} \approx 1.32$. Trajectories corresponding to $\xi=\xi_\text{f}$ are shown as black dashed lines in both (A) and (B).}
    \label{fig:evolution}
\end{figure}

Having derived analytical expressions for all the eigenstates of the system, we are ready to derive the dynamical properties of PDC. For an initial state $\ket{b_0}=\hat{v}^\dagger\ket{0}$, the state after some normalized time $\tau=\kappa t$ is 
\begin{align}
\label{eq:timeevolution}
    \ket{\Psi(\tau)}=&e^{\mathrm{i}\lambda_{\mathrm{M}}\tau}c_{\text{M}}\ket{\varphi_{\text{M}}}+\int_0^\infty\mathrm{d}\lambda \, c_{\lambda} e^{-\mathrm{i}\lambda \tau}\ket{\varphi_{\lambda}},
\end{align}
so, up to an overall phase, the pump state amplitude is
\begin{align}
\label{eq:ctau}
    C(\tau)=c_\mathrm{\text{M}}^2+\int_0^\infty\mathrm{d}\lambda \, c_{\lambda}^2 e^{-\mathrm{i}(\lambda+\lambda_\text{M}) \tau},
\end{align}
and the population of the pump photon $N_\mathrm{b}(\tau)=\left\vert C(\tau)\right\vert^2$ can be written as
\begin{small}\begin{align} \label{eq:evolution}
\!\!N_\mathrm{b}(\tau)=\left\lvert \left(1+\frac{\pi}{4\lambda_\mathrm{M}^{3/2}}\right)^{-1}+\int_0^\infty \!\mathrm{d}\lambda\, \frac{2\sqrt{\lambda}e^{-\mathrm{i}(\lambda+\lambda_\mathrm{M})\tau}}{4\lambda(\lambda-\xi)^2+\pi^2} \right\rvert^2.
\end{align}\end{small}In Fig.~\ref{fig:evolution}, we show the time evolution of $C(\tau)$ and $N_\mathrm{b}(\tau)$ for various values of $\xi$, both according to these analytic results as well as according to numerical simulation. We see that \eqref{eq:evolution} correctly predicts the dynamics under the discrete Hamiltonian \eqref{eq:ghamiltonian} for $\epsilon\ll 1$.  Interestingly, as shown by the dashed gray curve in Fig.~\ref{fig:evolution}, there exists a detuning $\xi = \xi_\text{f} \approx \num{1.90}$ which attains perfect pump depletion at a finite time $\tau = \tau_\text{f} \approx \num{1.32}$.

The analytic form of \eqref{eq:evolution} provides interesting insight into the dynamics of pump photon number. Since the continuum contribution $c_{\lambda}^2$ has finite support, its Fourier transform in \eqref{eq:evolution} should decay to zero as $\tau \rightarrow \infty$, so the pump photon population is eventually dominated by the meson contribution, i.e. $\lim_{\tau\rightarrow\infty}N_\mathrm{b}(\tau)=c_\mathrm{\text{M}}^2$. We can now consider two limits for the detuning $\xi$.

First, for large positive $\xi$, the continuum contribution $c_{\lambda}^2$ has a Lorentzian-like form with full-width-at-half-maximum of $\pi/\sqrt{\xi}$ near its peak $\lambda\approx\xi$, and thus, the continuum contribution decays to zero with characteristic time scale $\tau_\mathrm{d}=\sqrt{\xi}/\pi$, as argued above. In addition, for $\xi\rightarrow\infty$, the residual meson contribution has an asymptotic scaling $\mathrm{c}_{\text{M}}^2\sim \frac{\pi^2}{2\xi^3}$, so by simply taking large enough $\xi$ the residual pump population can be made arbitrarily small, at the cost of realizing this near-complete pump depletion at time $\tau_\mathrm{d} \propto \sqrt{\xi}$.

\begin{figure}[t]
\centering
    \includegraphics[width=0.48\textwidth]{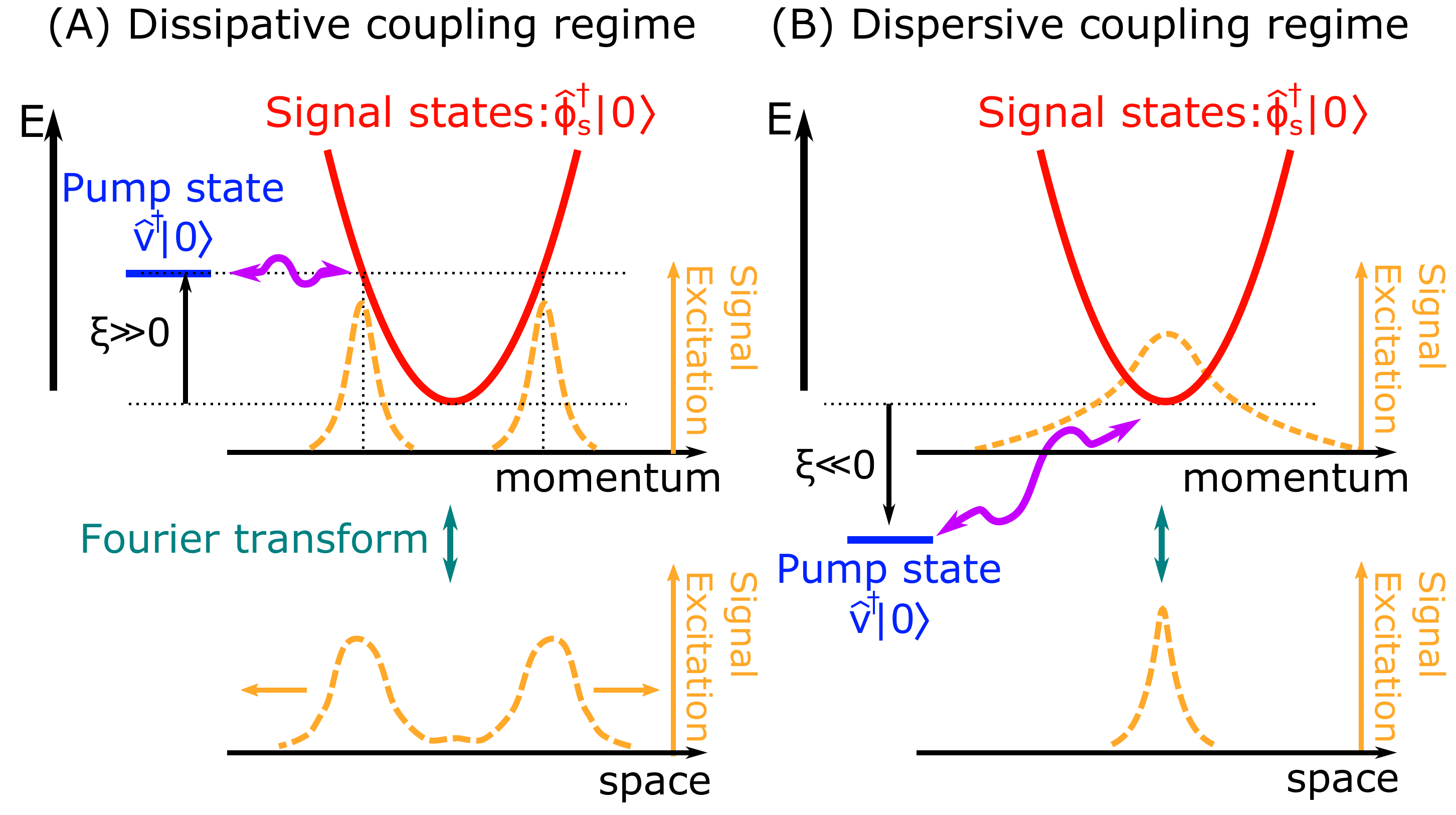}
    \caption{Illustration of single-photon PDC in (A) the dissipative coupling regime where the normalized detuning (i.e., phase mismatch) $\xi\rightarrow\infty$, and (B) the dispersive coupling regime where $\xi\rightarrow-\infty$. In the dissipative coupling regime (A), downconversion mostly populates signal states with energy close to that of the initial pump state, resulting in a Lorentzian lineshape in the continuum excitation; in the spatial domain, such an excitation corresponds to two wavepackets with opposite group velocity moving away from each other. In the dispersive coupling regime (B), the far-negatively-detuned initial pump state excites signal states near the bottom of the energy band upon downconversion, resulting in a spatially localized structure composed mostly of the optical meson solution.}
    \label{fig:dispersive_dissipative}
\end{figure}

As shown in Fig.~\ref{fig:dispersive_dissipative}(A), in the limit $\xi\rightarrow\infty$, the energy of the pump state lies deep in the middle of the signal continuum. In this regime (as shown in Sec.~\ref{sec:spatialcorrelation}), the downconverted signal photons populate wavepackets with large opposite group velocity, resulting in the suppression of backconversion. In other words, excitations that move from the pump state into the signal continuum do not come back into the pump, and the pump photon population exhibits an exponential decay
\begin{align}
    N_\text{b}(\tau)\approx e^{-\tau/\tau_\text{d}} \quad (\xi\rightarrow\infty).
\end{align}
We refer to this regime as dissipative single-photon PDC since the process results in the ``dissipation'' of a pump photon into the signal continuum. Additional details, analytics, and numerical results to support this interpretation are presented in Appendix~\ref{sec:asymptotic}.

On the other hand, in the other limit $\xi\rightarrow -\infty$, the energy of the pump state is far below the bottom of the signal continuum, so the pump state is dispersively coupled to a narrow band of signal states at the bottom of the signal continuum, as schematically shown in Fig.~\ref{fig:dispersive_dissipative}(B). The first consequence of this dispersive coupling is that downconverted signal photons tend to have small momenta and hence have wavepackets that separate slowly (as shown in Sec.~\ref{sec:spatialcorrelation}), allowing for multiple periods of Rabi-like oscillations in the pump photon number due to backconversion. At the same time, in contrast to the case of $\xi\rightarrow\infty$, the continuum contribution itself becomes small in this limit, with the dynamics being dominated by the static meson contribution $c_\text{M}^2 \rightarrow 1$ as $\xi \rightarrow-\infty$. More precisely, as shown in Appendix~\ref{sec:asymptotic}, the pump population as a function of time can be approximated as
\begin{small}\begin{align}
    \label{eq:negative_osccilation}
    N_\text{b}(\tau)\approx \left\vert 1-\frac{\pi}{4(-\xi)^{3/2}}+\frac{\sqrt{\pi}e^{\mathrm{i}\left(\xi \tau-\frac{\pi}{4}\right)}}{2\xi^2\sqrt{\tau}}\right\vert^2 \,\, (\xi\rightarrow-\infty),
\end{align}\end{small}which exhibits the aforementioned oscillations (third term) on top of the static meson contribution (first two terms). Notably, this expression indicates that the oscillations follow a sub-exponential decay $\sim 1/\sqrt{\tau}$.

It is worth mentioning that, even in the dispersive limit $\xi\rightarrow-\infty$, \eqref{eq:negative_osccilation} exhibits qualitatively different dynamics from conventional Rabi oscillations seen in \emph{single-mode} PDC with Hamiltonian~\cite{Drobny1994,Bandilla2000}
\begin{align}
    \label{eq:singlemodehamiltonian}
    \frac{g_\text{s}}{2}\Bigl(\hat{a}_\text{s}^2\hat{b}_\text{s}^\dagger+\hat{a}_\text{s}^{\dagger 2}\hat{b}_\text{s}\Bigr)+\delta_s\hat{a}_\text{s}^\dagger\hat{a}_\text{s}.
\end{align}
For an initial single-photon pump state $\hat{b}_\text{s}^\dagger\ket{0}$, \eqref{eq:singlemodehamiltonian} exhibits a sinusoidal oscillation with no decay: the pump photon population exactly returns to unity after a period of oscillation regardless of the detuning $\delta_\text{s}$. On the other hand, after one oscillation period, \eqref{eq:negative_osccilation} attains a diminished peak value of the pump population
\begin{align}
    N_\text{b}\left(\tau=-\frac{7\pi}{4\xi}\right)\approx 1-\left(\frac{\pi}{2}-\frac{2}{\sqrt{7}}\right)(-\xi)^{-3/2},
\end{align}
which is smaller by $\sim(-\xi)^{-3/2}$. But because the amplitude of the oscillation is also $\sim(-\xi)^{-3/2}$, \eqref{eq:negative_osccilation} cannot recover the Rabi oscillations of \eqref{eq:singlemodehamiltonian} even in the limit $\xi \rightarrow-\infty$. The fast initial decay induced by the characteristic scaling $\sim 1/\sqrt{\tau}$ indicates that the broadband PDC is inherently multimode even in the dispersive limit where the coupling of the pump to the signal continuum might be thought of as being ``narrowband''.

\subsection{Signal biphoton correlation functions}
\label{sec:spatialcorrelation}
\begin{figure*}[hbt]
\centering
    \includegraphics[width=0.95\textwidth]{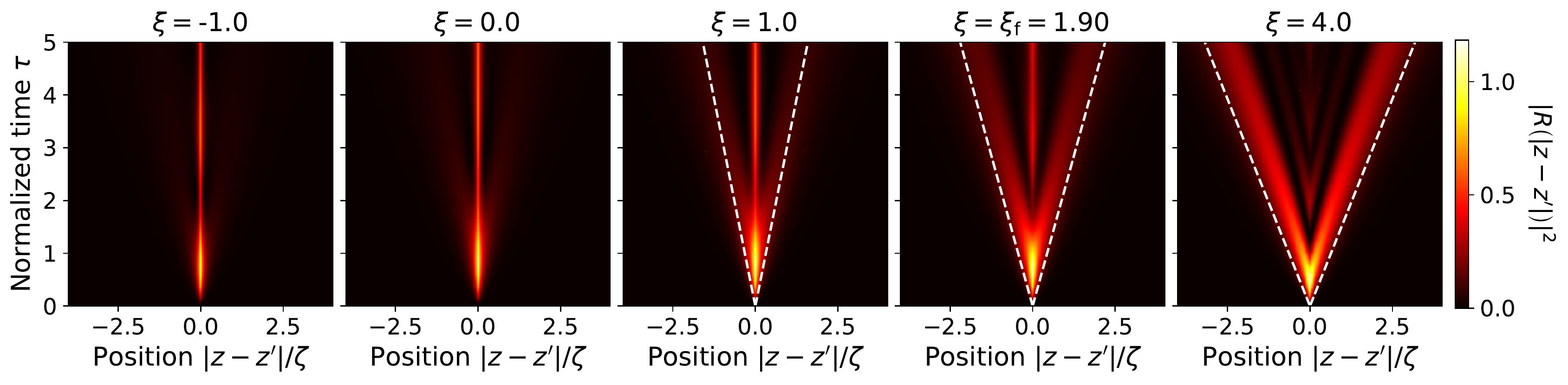}
    \caption{Evolution of the amplitude of the spatial biphoton correlation function $|R(|z-z'|)|^2$ for single-photon PDC under various normalized detunings (i.e., phase mismatch) $\xi$, using the analytic continuum expression given by \eqref{eq:temporal}. For negative $\xi$, the dynamics mostly excite the meson state, resulting in spatially localized signal photons throughout propagation. For large positive $\xi$, the downconverted signal photons populate states with finite momentum, resulting in two counterpropagating wavepackets with group velocity $\pm\sqrt{\xi}\zeta/\pi$ (indicated by white dashed lines as a guide). For an intermediate $\xi$, signal photons populate both the meson state and finite-momentum continuum states, resulting in triplet peaks. Here, $\xi_\text{f} \approx 1.90$ refers to the particular detuning for finite-time pump depletion discussed in Fig.~\ref{fig:evolution}.}
    \label{fig:temporal}
\end{figure*}

Because there is only one pump state involved in the single-photon PDC dynamics (i.e., the dc pump state), $N_\mathrm{b}(\tau)$ suffices to capture the information about the pump population dynamics. On the other hand, the downconverted signal pair can populate various momentum modes, so a full characterization of the PDC process also requires understanding the signal momentum distribution. In this section, we derive analytic expressions for the signal biphoton correlation function in the continuum approximation.

A general superposition state of a pump photon in a discrete dc mode $\hat b$ and two signal photons somewhere within the quantization window can be parameterized as
\begin{align}
\label{eq:temporalcorrelation}
    \left(C\hat{b}_0^\dagger+\frac{1}{\sqrt{L}}\iint^{L/2}_{-L/2}\mathrm{d}z\,\mathrm{d}z' \, R(z,z')\hat{\psi}_{z}^\dagger\hat{\psi}_{z'}^\dagger\right)\ket{0},
\end{align}
where $|C|^2$ is the pump population and $R(z,z')$ is the biphoton correlation function. Here, $\hat{\psi}_z$ are local signal field annihilation operators with commutation relationships $[\hat{\psi}_z,\hat{\psi}_{z'}^\dagger]=\delta(z-z')$, related to $\hat{a}_m$ via
\begin{align}
    \hat{a}_m=\frac{1}{\sqrt{L}}\int^{L/2}_{-L/2}\mathrm{d}z \, e^{2\pi\mathrm{i} mz/L}\hat{\psi}_z.
\end{align}
Note that due to the commutation relationships for $\hat\psi_z$ arising from particle indistinguishability, the biphoton correlation function is symmetric, i.e., $R(z,z')=R(z',z)$. In the following, we calculate $R(z,z')$ in the continuum limit $L\rightarrow\infty$ using the results obtained in the previous section.

We first express the time evolution of the state $\ket{\Psi(\tau)}$ in terms of the $\hat b_0$ and the continuum signal-pair momentum operators $\hat\phi_s$ by substituting \eqref{eq:eigen} into \eqref{eq:timeevolution}:
\begin{subequations}
\begin{align} \label{eq:spectralcorrelation}
\ket{\Psi(\tau)}=\left(C(\tau)e^{\mathrm{i}\lambda_\text{M}\tau}\hat{b}_0^\dagger+\int^\infty_0\mathrm{d}s \, Q(\tau,s) \hat\phi^\dagger_s\right)\ket{0}
\end{align}
where
\begin{align}
Q(\tau,s)=&c_{\text{M}}f_{\text{M}}(s)e^{\mathrm{i}\lambda_\text{M}\tau}+\int^\infty_0\mathrm{d}\lambda \, c_{\lambda} f_{\lambda}(s)e^{-\mathrm{i}\lambda\tau}
\end{align}
\end{subequations}
gives the amplitude for photon pairs with non-degeneracy $s$ (related to their wavevectors by $k=\pm 2\pi s/\epsilon L$). Here, $Q(\tau,s)$ can be seen as a spectral biphoton correlation function and, intuitively, is related to the spatial biphoton correlation function $R(z,z')$ via a Fourier transformation as follows. First, to relate $\hat \phi_s$ to $\hat\psi_z$, we can, in the continuum limit of large $L$, make the formal substitution
\begin{align}
    \!\hat{\phi}_s&\mapsto\frac{1}{\epsilon^{\frac{1}{2}}L}\iint^{L/2}_{-L/2}\mathrm{d}z\mathrm{d}z' \, \exp\bigl(2\pi\mathrm{i} s(z-z')/\epsilon L\bigr)\hat{\psi}_z\hat{\psi}_{z'}\nonumber\\
    &=\frac{1}{\sqrt{\zeta L}}\iint^{L/2}_{-L/2}\mathrm{d}z\mathrm{d}z'\cos\bigl(2\pi s(z-z')/\zeta\bigr)\hat{\psi}_z\hat{\psi}_{z'},
\end{align}
where $\zeta=\epsilon L$ is a characteristic correlation length of the signal photons independent of $L$. Then, by comparing terms in \eqref{eq:temporalcorrelation} to the signal component of \eqref{eq:spectralcorrelation}, we derive
\begin{align}
    \label{eq:temporal}
    \!R(&z,z')=\frac{1}{\sqrt{\zeta}}\int^\infty_0 \mathrm{d}s \, Q(\tau,s)\cos\left(2\pi s (z-z')/\zeta\right)\nonumber\\
    =&-\frac{1}{\sqrt{\zeta}}\frac{2\pi\lambda_\mathrm{M}\exp\left(-2\pi\sqrt{\lambda_\mathrm{M}} |z-z'|/\zeta\right)e^{\mathrm{i}\lambda_\mathrm{M}\tau}}{\pi+4\lambda_\mathrm{M}^{3/2}}\\
    &+\frac{1}{\sqrt{\zeta}}\int^\infty_0\mathrm{d}\lambda \, \frac{\cos\bigl(2\pi\sqrt{\lambda} |z-z'|/\zeta+\Delta(\lambda)\bigr)e^{-\mathrm{i}\lambda \tau}}{\sqrt{w^2(\lambda)+\pi^2}},\nonumber
\end{align}
where $\Delta(\lambda)=-\arctan\bigl(\pi/w(\lambda)\bigr)$ is Fano's phase parameter. Here, the first term of Eq.~\eqref{eq:temporal} corresponds to the meson contribution, while the second term comes from the continuum states. Due to the transitionally invariant nature of the state $\ket{\Psi(\tau)}$, we note that $R(z,z')$ is a function of only $|z-z'|$; thus, we denote $R(z,z')=R(|z-z'|)$ in the following.

One way to intuitively understand the biphoton correlation function $R(|z-z'|)$ is to suppose one signal photon has been found at position $z'$. This projects the joint state $\ket{\Psi(\tau)}$ onto $\int\mathrm{d}z \, R(|z-z'|)\hat{\psi}_z\ket{0}$, up to normalization. Thus, $R(|z-z'|)$ corresponds to the spatial wavefunction of the second signal photon as a function of $z$, after finding the first signal photon to be located at $z'$.

In Fig.~\ref{fig:temporal}, we show the amplitude of the biphoton correlation function $|R(|z-z'|)|^2$ calculated based on \eqref{eq:temporal}. As discussed in the previous subsection and in Fig.~\ref{fig:dispersive_dissipative}, when $\xi$ is negative, most of the signal photons populate meson state, and they remain spatially localized over time. On the other hand, for large positive $\xi$, signal photons populate continuum modes around the energy $\lambda\sim\xi$. As a result, these signal photons with opposite momentum move away from each other, and back conversion is suppressed, which is exhibited in Fig.~\ref{fig:temporal} as two wavepackets moving apart. More precisely, since these signal photons predominantly have energy $\lambda \sim \xi$, their non-degeneracy $s \sim \sqrt{\lambda}$, which means their physical wavevectors are $k = \pm 2\pi s/\epsilon L \sim \pm 2\pi \sqrt{\lambda}/\zeta$. Thus, the group velocities of these wavepackets are approximately
\begin{align}
\left.\frac{\partial\lambda}{\partial k}\right\vert_{\lambda=\xi}=\left.\frac{\partial \lambda}{\partial \bigl(\pm 2\pi \sqrt{\lambda}/\zeta\bigr)}\right|_{\lambda=\xi}=\pm\frac{\sqrt{\xi}}{\pi}\zeta.
\end{align}
We see from Fig.~\ref{fig:temporal} that this approximation predicts well the observed spreading of the wavepackets.

This analysis of the biphoton correlation function has important implications for the characteristic correlation length $L_\text{c}$ discussed in Sec.~\ref{sec:model}. When choosing the quantization window length $L$, we require $L > L_\text{c}$ in order to fully capture non-local correlations out to a length $L_\text{c}$. We now see that this length is essentially set by the support of the biphoton correlation function $R(|z-z'|)$. In particular, the requirement is more stringent in the case of $\xi > 0$, where spatial spreading of the signal correlations are important; in this case, the above analysis shows that $L_\text{c}\sim \frac{2}{\pi}\sqrt{\xi}\zeta\tau$, with $\zeta$ being a characteristic scale for the growth of the signal-photon correlation length.

\section{PDC in coupled broadband nonlinear waveguides} \label{sec:coupled}
To demonstrate the utility of the formalism presented in Sec.~\ref{sec:singlephotonPDC}, we use the same approach to study Fano-type interactions in a model system composed of two $\chi^{(2)}$ waveguides with linear coupling between their continuum signal modes. In the weak-excitation limit, if we consider as input a single photon instantiated in a superposition state between the two waveguides, the simultaneous PDC of the discrete pump states of the two waveguides into the signal continuum can give rise to Fano interference in the quantum dynamics. This interference generates characteristically asymmetric Fano lineshapes in the continuum spectrum, and, under appropriate conditions, a sharp resonance can form within the spectrum that critically affects the PDC rate and also heralds the existence of a bound state in the continuum (BIC). This set of dramatic PDC dynamics, obtained by simply varying the input state and the relative waveguide detunings, are exactly analogous to the dynamics studied in other engineered systems exhibiting Fano interference, e.g., as demonstrated in Ref.~\cite{Crespi2015}.

\begin{figure*}[hbt]
    \centering
    \includegraphics[width=1.0\textwidth]{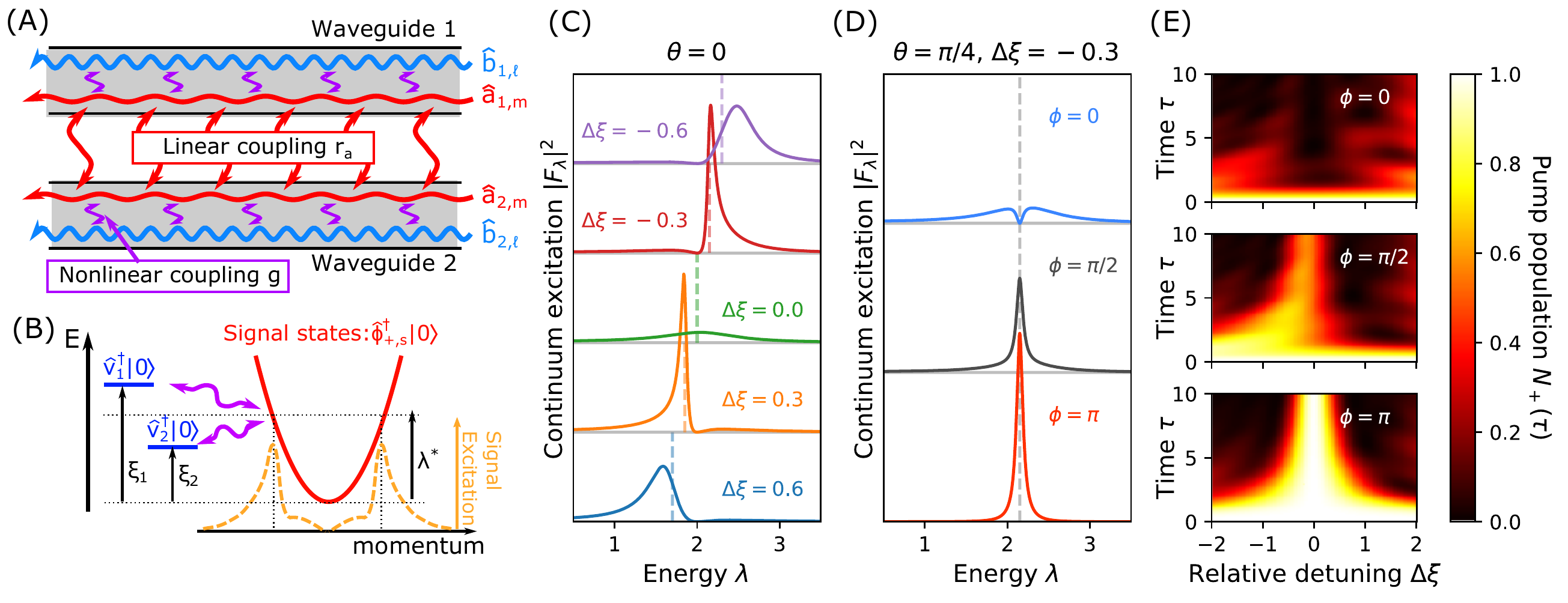}
    \caption{(A) Illustration of a pair of linearly coupled nonlinear waveguides. Signal modes $\hat{a}_{1,m}$ and $\hat{a}_{2,m}$ of waveguides 1 and 2, respectively, are linearly coupled with rate $r_\text{a}$. Independently, these modes nonlinearly interact (e.g., via a $\chi^{(2)}$ process) with each waveguide's respective pump modes $\hat{b}_{i,\ell}$ at rate $g$. (All other waveguide dispersion parameters are assumed identical.) (B) Illustration of the energy levels of the coupled-waveguide system in the continuum limit. Although the waveguides have different normalized detunings $\xi_1$ and $\xi_2$ (i.e., phase mismatches) in general, single-photon pump states of the two waveguides (denoted $\hat{v}_1^\dagger\ket{0}$ and $\hat{v}_2^\dagger\ket{0}$) can nevertheless downconvert into a common signal continuum $\hat{\phi}_{+,s}^\dagger\ket{0}$ (where ``+'' indicates, e.g., the symmetric spatial mode of the waveguides). The resulting signal excitation (dashed orange curve) consists of PDC from both pump states and can therefore support Fano interference between the two processes. (C) Continuum (eigenstate) excitation $|F_\lambda|^2$ according to \eqref{eq:f_lambda} with various relative detunings $\Delta\xi=\xi_2-\xi_1$, for an input state \eqref{eq:initial_coupled} with $\theta=0$ (i.e., single pump photon input to waveguide 1). Vertical dashed lines indicate $\lambda=\lambda^*=\frac12(\xi_1+\xi_2)$, around which the spectra show characteristic Fano lineshape asymmetries (e.g., peak to one side and zero to the other side). At $\Delta\xi=0$ (green curve), the continuum resonance peak becomes infinitely sharp and the system supports a bound state in the continuum (BIC), which is not shown. (D) Plots of $|F_\lambda|^2$ for input states with $\theta=\pi/4$ (i.e., single pump photon initialized in an equal superposition between waveguides 1 and 2) and relative phase $\phi$, but with fixed relative detuning $\Delta\xi=-0.3$. For $\phi=0$ ($\phi=\pi$), the two PDC processes constructively (destructively) interfere, resulting in faster (slower) downconversion accompanied by wider (narrower) continuum lineshapes. (E) Evolution of the pump photon population $N_+(\tau)$ for various $\Delta\xi$ and the same set of input states as in (D), showing the expected variation in PDC rates. At $\Delta\xi=0$, the input state with $\phi=\pi$ directly excites the BIC and does not decay (i.e., downconvert). In all the above figures, we fix $\xi_2=2$, with $\xi_1$ set by $\Delta\xi$ accordingly.}
    \label{fig:fano_combined}
\end{figure*}

As shown in Fig.~\ref{fig:fano_combined}(A), we consider two waveguides 1 and 2 with (photon-polariton) field annhilation operators $\hat{a}_{i,m}$ for signal and $\hat{b}_{i,\ell}$ for pump, where $i\in\{1,2\}$ represents the index of the waveguides. We denote by $\delta_i$ the respective detunings (i.e., phase mismatch), which are not necessarily equal, although we assume for simplicity that the nonlinearity $g$, linear energy dispersion $\mu$, and quadratic energy dispersions $d_\text{a},d_\text{b}$ are identical between the two waveguides. Additionally, as in Sec.~\ref{sec:singlephotonPDC}, we assume $d_\text{a}>0$. The total Hamiltonian of the coupled waveguide system is
\begin{align} \label{eq:coupledhamiltonian}
    \hat{H}_1+\hat{H}_2+\hat{V},
\end{align}
where $\hat{H}_{i}$ is obtained by respectively replacing $\hat{a}_m$, $\hat{b}_\ell$, and $\delta$ with $\hat{a}_{i,m}$, $\hat{b}_{i,\ell}$ and $\delta_i$ in \eqref{eq:fullhamiltonian}. We assume the signal modes of the two waveguides are linearly coupled to each other (e.g., via evanescent fields), which we describe by an interaction term
\begin{align}
    \hat{V}/\hbar&=-r_\text{a}\sum_{m=-\infty}^\infty\left(\hat{a}_{1,m}^\dagger\hat{a}_{2,m}+\hat{a}_{1,m}\hat{a}_{2,m}^\dagger\right),
\end{align}
where $r_\text{a}>0$ is the strength of the coupling. Physically, $\hat{V}$ represents a linear coupling which is local in the spatial domain. Using coupled mode theory~\cite{Huang1994}, we can equivalently describe the system using symmetric and anti-symmetric transverse waveguide modes
\begin{align}
    \hat{a}_{\pm,m}=\frac{\hat{a}_{1,m}\pm\hat{a}_{2,m}}{\sqrt{2}},
\end{align}
which allows us to write the interaction as
\begin{align}
\label{eq:eshift}
    \hat{V}/\hbar=-r_\text{a}\sum_{m=-\infty}^\infty(\hat{a}_{+,m}^\dagger\hat{a}_{+,m}-\hat{a}_{-,m}^\dagger\hat{a}_{-,m});
\end{align}
i.e., the symmetric (anti-symmetric) signal modes experience a negative (positive) energy shift by $r_\text{a}$.

We now assume the waveguides are designed so that the energy shift caused by the coupling \eqref{eq:eshift} results in the anti-symmetric signal modes being far-detuned relative to the energy scale of the other dynamics, i.e., $\left|2r_\text{a}-\delta_i\right|\gg\kappa$, where the characteristic nonlinear rate $\kappa$ is defined by \eqref{eq:kappa} as usual. Under this assumption, the anti-symmetric signal modes $\hat{a}_{-,m}$ remain unpopulated throughout the system evolution, and we therefore drop terms involving these modes from the dynamics, leading to a reduced Hamiltonian
\begin{align}
        \begin{split}
        \hat{H}_{+}/\hbar&=\frac{g}{4}\sum_{m+n=\ell}\left(\hat{a}_{+,m}^\dagger \hat{a}_{+,n}^\dagger(\hat{b}_{1,\ell}+\hat{b}_{2,\ell})+\mathrm{H.c.}\right)\\
        &+\sum_{i=1,2}\sum_{\ell=-\infty}^\infty \left(\delta_i+2r_\text{a}+\mu\ell +d_\text{b}\ell^2\right)\hat{b}_{i,m}^\dagger\hat{b}_{i,m}\\
        &+\sum_{m=-\infty}^\infty d_\text{a}m^2\hat{a}_{+,m}^\dagger\hat{a}_{+,m},\\
    \end{split}
\end{align}
where we have moved into a new rotating frame via substitutions $\hat{a}_{+,m}\mapsto e^{-\mathrm{i} r_\text{a}t}\,\hat{a}_{+,m}$ and $\hat{b}_{i,\ell}\mapsto e^{- 2\mathrm{i}r_\text{a}t}\,\hat{b}_{i,\ell}$.

For this system, we consider the PDC of an initial monochromatic ($\ell = 0$) single-photon state instantiated in the superposition
\begin{align}
\label{eq:initial_coupled}
    \ket{\Psi_+(\tau=0)}=\left(\cos\theta\, \hat{b}_{1,0}^\dagger+e^{\mathrm{i}\phi}\sin\theta\,\hat{b}_{2,0}^\dagger\right)\ket{0},
\end{align}
for which the only states relevant to the dynamics are
\begin{subequations}
    \begin{align}
        \ket{b_{i,0}}&=\hat{b}_{i,0}^\dagger\ket{0}\\
            \ket{a_{+,p}} &= \begin{cases}
                \frac{1}{\sqrt{2}}\hat{a}_{+,0}^{\dagger2}\ket{0} & p=0 \\
                \hat{a}^\dagger_{+,p}\hat{a}_{+,-p}^\dagger\ket{0} & \text{otherwise}
            \end{cases}.
    \end{align}
\end{subequations}
Introducing $\hat{G}_+=\hat{H}_+/\hbar\kappa$ to be the normalized Hamiltonian as in Sec.~\ref{sec:singlephotonPDC}, its non-zero matrix elements among these states are
\begin{subequations} 
    \begin{align}
        \braket{b_{i,0}|\hat{G}_+|b_{i,0}} &= \xi_i \\
        \braket{a_{+,p}|\hat{G}_+|a_{+,p}} &= \epsilon^2p^2 \\
        \braket{a_{+,p}|\hat{G}_+|b_{i,0}}& = \frac{1}{2}\epsilon^{1/2},
    \end{align}
\end{subequations}
where $\xi_i=(\delta_i+2r_\text{a})/\kappa$ are the normalized detunings. Following the same procedure as in Sec.~\ref{sec:singlephotonPDC}, we obtain a continuum Hamiltonian $\hat{G}^\text{cont}_+=\lim_{L\rightarrow\infty}\hat{G}_+$ as
\begin{subequations}
\begin{align}
    \label{eq:continuum_coupled}
    \!\hat{G}^\text{cont}_+=& \sum_{i=1,2}\left[\xi_i\hat{v}_i^\dagger\hat{v}_i+\frac{1}{2}\int^\infty_0\mathrm{d}s\left(\hat{v}_i^\dagger\hat{\phi}_{+,s}+\hat{v}_i\hat{\phi}_{+,s}^\dagger\right)\right]\nonumber\\
    &{}+\int^\infty_0\mathrm{d}s\,s^2\hat{\phi}_{+,s}^\dagger \hat{\phi}_{+,s},
\end{align}
where
\begin{align}
    &\hat{v}_i=\ket{0}\bra{b_{i,0}}, &\hat{\phi}_{+,s}=\lim_{L\rightarrow\infty}\epsilon^{-\frac{1}{2}}\ket{0}\bra{a_{+,p}}
\end{align}
\end{subequations}
with $s=\epsilon p$. The energy levels of the continuum Hamiltonian \eqref{eq:continuum_coupled} are schematically shown in Fig.~\ref{fig:fano_combined}(B).

Again using Fano's theory for discrete-continuum interaction in Ref.~\cite{Fano1961}, we assume the eigenstates of $\hat{G}^\text{cont}_+$ with eigenvalue $\lambda$ take the form
\begin{align}
    \ket{\varphi_{+,\lambda}}=\Biggl(\,\sum_{i=1,2} c_{i,\lambda}\hat{v}_i^\dagger+\int_0^\infty \mathrm{d}s \, f_{+,\lambda}(s)\hat{\phi}_{+,s}^\dagger\Biggr)\ket{0},
\end{align}
which leads to a set of equations
\begin{subequations}
    \begin{align}
        \label{eq:c12eq}
        \xi_ic_{i,\lambda}+\frac{1}{2}\int^\infty_0\mathrm{d}s \, f_{+,\lambda}(s)&=\lambda c_{i,\lambda}\quad(i=1,2)\\
        \label{eq:feq}
        s^2f_{+,\lambda}(s)+\frac{1}{2}\sum_{i=1,2}c_{i,\lambda}&=\lambda f_{+,\lambda}(s).
    \end{align}
\end{subequations}

As before, we first consider solutions with negative energy $\lambda=-\lambda_\text{B}<0$, where the subscript ``B'' stands for bound states. We denote $c_{i,\text{B}}=c_{i,-\lambda_\text{B}}$, $f_{+,\text{B}}=f_{+,-\lambda_\text{B}}$, and $\ket{\varphi_{+,\text{B}}}=\ket{\varphi_{+,-\lambda_\text{B}}}$. For the signal part, the solution to \eqref{eq:feq} is
\begin{align}
    f_{+,\text{B}}(s)=-\frac{c_{1,\text{B}}+c_{2,\text{B}}}{2(\lambda_\text{B}+s^2)},
\end{align}
and substituting this result into \eqref{eq:c12eq} yields
\begin{align}
    \label{eq:series_bound}
    \begin{pmatrix}
        \xi_1+\lambda_\text{B}-\frac{\pi}{8\sqrt{\lambda_\text{B}}} & -\frac{\pi}{8\sqrt{\lambda_\text{B}}}\\
        -\frac{\pi}{8\sqrt{\lambda_\text{B}}}&\xi_2+\lambda_\text{B}-\frac{\pi}{8\sqrt{\lambda_\text{B}}}
    \end{pmatrix}\begin{pmatrix}
        c_{1,\text{B}}\\
        c_{2,\text{B}}
    \end{pmatrix}=0.
\end{align}
Nontrivial solutions to \eqref{eq:series_bound} exist when the bound state energies $\lambda_\text{B}$ solve the secular equation
\begin{align}
    \!\!\frac{\pi}{8\sqrt{\lambda_\text{B}}}(\xi_1+\xi_2)+\frac{\pi\sqrt{\lambda_\text{B}}}{4}-(\lambda_\text{B}+\xi_1)(\lambda_\text{B}+\xi_2)=0,
\end{align}
which has two positive solutions for $\xi_1+\xi_2<0$ and a single solution for $\xi_1+\xi_2\geq0$. In the former case ($\xi_1+\xi_2<0$), we have two bound states outside of the signal energy band, which are analogous to the optical meson states we obtained in Sec.~\ref{sec:singlephotonPDC}. On the other hand, for the latter case ($\xi_1+\xi_2\geq0$), there exists only one optical meson-like bound state below the signal continuum.

Normalization of these bound states requires
\begin{align}
\label{eq:normalization}
    c_{1,\text{B}}^2+c_{2,\text{B}}^2+\frac{\pi}{16\lambda_\text{B}^{3/2}}(c_{1,\text{B}}+c_{2,\text{B}})^2=1.
\end{align}
From \eqref{eq:series_bound}, we have $c_{1,\text{B}}(\lambda_\text{B}+\xi_1)=c_{2,\text{B}}(\lambda_\text{B}+\xi_2)$, meaning we can write
\begin{align}
    c_{1,\text{B}}=S_\text{B}^{-\frac{1}{2}}(\lambda_\text{B}+\xi_2), \quad c_{2,\text{B}}=S_\text{B}^{-\frac{1}{2}}(\lambda_\text{B}+\xi_1),
\end{align}
where using \eqref{eq:normalization},
\begin{align}
    S_\text{B} =(\lambda_\text{B}+\xi_1)^2+(\lambda_\text{B}+\xi_2)^2 +\frac{\pi}{16\lambda_\text{B}^{3/2}}(2\lambda_\text{B}+\xi_1+\xi_2)^2.\nonumber
\end{align}
These equations provide complete analytic expressions for the bound state solutions with energy $\lambda=-\lambda_\text{B}<0$.

Next, we consider solutions with non-negative energies $\lambda\geq 0$. As in Sec.~\ref{sec:singlephotonPDC}, we write such solutions to \eqref{eq:feq} in the form
\begin{align}
    \label{eq:fplus}
    f_{+,\lambda}=&\frac{c_{1,\lambda}+c_{2,\lambda}}{2}\left(\frac{1}{\lambda-s^2}+w_{+}(\lambda)\delta(\lambda-s^2)\right),
\end{align}
where $w_{+}(\lambda)$ is to be determined by the other conditions. More specifically, a substitution of \eqref{eq:fplus} into \eqref{eq:c12eq} yields
\begin{align}
    \label{eq:series}
    \begin{pmatrix}
        \xi_1-\lambda+\frac{w_{+}(\lambda)}{8\sqrt{\lambda}}&\frac{w_{+}(\lambda)}{8\sqrt{\lambda}}\\
        \frac{w_{+}(\lambda)}{8\sqrt{\lambda}}&\xi_2-\lambda+\frac{w_{+}(\lambda)}{8\sqrt{\lambda}}
    \end{pmatrix}\begin{pmatrix}
        c_{1,\lambda}\\
        c_{2,\lambda}
    \end{pmatrix}=0.
\end{align}
Nontrivial solutions to \eqref{eq:series} exist for $\lambda$ which solve the secular equation 
\begin{align}
    \!\!\bigl(2\lambda-(\xi_1+\xi_2)\bigr)w_{+}(\lambda)-8\sqrt{\lambda}(\lambda-\xi_1)(\lambda-\xi_2)=0,
\end{align}
which, for $\lambda\neq \frac12(\xi_1+\xi_2)$, has a solution
\begin{align}
    w_{+}(\lambda)=\frac{8\sqrt{\lambda}(\lambda-\xi_1)(\lambda-\xi_2)}{2\lambda-(\xi_1+\xi_2)}.
\end{align}
Normalization of these continuum states requires
\begin{align}
\label{eq:normalization_cont}
    \left(c_{1,\lambda}+c_{2,\lambda}\right)^2=\frac{8\sqrt{\lambda}}{\pi^2+w_{+}^2(\lambda)}.
\end{align}
From \eqref{eq:series} we have $c_{1,\lambda}(\lambda-\xi_1)=c_{2,\lambda}(\lambda-\xi_2)$, meaning we can write
\begin{align} \label{eq:c12_cont}
    c_{1,\lambda}=S_\lambda^{-\frac{1}{2}}(\lambda-\xi_2),\quad c_{2,\lambda}=S_\lambda^{-\frac{1}{2}}(\lambda-\xi_1),
\end{align}
where using \eqref{eq:normalization_cont},
\begin{align*}
S_\lambda=\frac{\pi^2+w_+^2(\lambda)}{8\sqrt{\lambda}}\Bigl(2\lambda-(\xi_1+\xi_2)\Bigr)^2.
\end{align*}

It is notable that the continuum has a ``hole'' at the energy $\lambda^*=\frac12(\xi_1+\xi_2)$, and generally, there is no continuum solution at this point. However, by examining the expression for \eqref{eq:c12_cont}, we see that $c_{i,\lambda}$ is sharply peaked in the vicinity of $\lambda^*$, indicating the presence of a resonance in the continuum; as we later show numerically, this resonance becomes sharper as the relative detuning $\Delta\xi=\xi_2-\xi_1$ approaches zero. Analytically, there appears a special bound state at $\Delta\xi=0$, where \eqref{eq:fplus} and \eqref{eq:series} support a nontrivial solution $c_1=-c_2=\frac{1}{\sqrt{2}}$ and $f_{+,\lambda^*}=0$, i.e., consisting only of pump photons. Because the energy $\lambda^*\geq0$ of this bound state lies in the continuum, it can be seen as a bound state in the continuum (BIC)~\cite{Hsu2016}. Such appearance of a BIC at a high-symmetry point has been shown in Ref.~\cite{Crespi2015}, for instance. In the following, we focus on the case $\xi_1+\xi_2\geq0$ to further investigate the exotic dynamics of discrete-continuum interferences especially in the vicinity of BIC.

Having derived the complete eigenspectrum of the system given by \eqref{eq:continuum_coupled}, we now consider its dynamical properties. For the initial state $\ket{\Psi_+(\tau=0)}$ given by \eqref{eq:initial_coupled}, its time evolution after a time $\tau$ is
\begin{subequations}
\begin{align}
    \label{eq:stateevolution}
    \begin{split}
    \ket{\Psi_+(\tau)}=&(\cos\theta c_{1,\text{B}}+e^{\mathrm{i}\phi}\sin\theta c_{2,\text{B}})e^{\mathrm{i}\lambda_\text{B}\tau}\ket{\varphi_{+,\text{B}}}\\
    &+\int^\infty_0\mathrm{d}\lambda \,  F_\lambda(\theta,\phi) e^{-\mathrm{i}\lambda\tau}\ket{\varphi_{+,\lambda}},
    \end{split}
\end{align}
where
\begin{align}
\label{eq:f_lambda}
    F_\lambda(\theta,\phi)=\cos\theta c_{1,\lambda}+e^{\mathrm{i}\phi}\sin\theta c_{2,\lambda}
\end{align}
\end{subequations}
represents the excitation amplitude spectrum of the initial state. In Fig.~\ref{fig:fano_combined}(C), we show $|F_\lambda|^2$ for $\theta=0$ (i.e., single-photon input to waveguide 1) at different relative detunings $\Delta\xi=\xi_2-\xi_1$. We observe characteristic asymmetric Fano resonance lineshapes, critically dependent on $\Delta\xi$. As a function of $\lambda$, the spectrum shows a resonance peak around $\lambda^*$, and the resonance becomes sharper as $\Delta\xi\rightarrow0$. At $\Delta\xi=0$, the peak becomes infinitely narrow to form a BIC at energy $\lambda=\lambda^*$. (However, as the BIC solution is not included in the expression of $F_\lambda$, we see no peak in the figure at $\Delta\xi=0$.)

We can also study the effects of interference due to non-local correlations in the initial pump state between the two waveguides. More specifically, we set $\theta=\frac{\pi}{4}$ and vary $\phi$, i.e., $\ket{\Psi_+(0)}=\frac{1}{\sqrt{2}}\bigl(\hat{v}_1^\dagger+e^{\mathrm{i}\phi}\hat{v}_2^\dagger\bigr)\ket{0}$. As shown by plots for $|F_\lambda|^2$ in Fig.~\ref{fig:fano_combined}(D), the width of the resonance peaks change dramatically as a function of $\phi$. For $\phi=0$, the couplings between the two pump states and the signal continuum constructively interfere, leading to a faster decay (i.e., PDC rate), accompanied by a broader resonance peak. On the other hand, for $\phi=\pi$, these two paths destructively interfere and suppress the rate of PDC, exhibiting a sharp resonance which increases the lifetime of the pump photons.

We can equivalently see this modulation of the PDC rate by computing the dynamics of the pump photon population as in Sec.~\ref{sec:singlephotonPDC}. An analytic expression for the pump photon population $N_+(\tau)$ follows from \eqref{eq:stateevolution} as
\begin{subequations}
\begin{align}
    \label{eq:coupled_pump}
N_+(\tau)=|C_1(\tau)|^2+|C_2(\tau)|^2
\end{align}
where
\begin{align}
C_i(\tau)&=c_{i,\text{B}}\left(\cos\theta c_{1,\text{B}}+e^{\mathrm{i}\phi}\sin\theta c_{2,\text{B}}\right)e^{\mathrm{i}\lambda_\text{B}\tau}\\
    &\,\,{}+\int^\infty_0\mathrm{d}\lambda\, c_{i,\lambda}\left(\cos\theta c_{1,\lambda}+e^{\mathrm{i}\phi}\sin\theta c_{2,\lambda}\right)e^{-\mathrm{i}\lambda\tau}.\nonumber
\end{align}
\end{subequations}
We show the evolution of $N_+(\tau)$ for various $\Delta\xi$ and $\phi$ with $\theta=\frac{\pi}{4}$ using \eqref{eq:coupled_pump} in Fig.~\ref{fig:fano_combined}(E). As expected from the arguments above, the conversion rate is enhanced for $\phi=0$, while the lifetime of the pump photon is significantly longer for $\phi=\pi$. In particular, at $\Delta\xi=0$, an input state with $\phi=\pi$ directly populates only the BIC, and as a result, no PDC occurs at all (i.e., $N_+(\tau)=1$) despite the presence of $\chi^{(2)}$ nonlinear interactions. For an intermediate $\phi=\frac{\pi}{2}$, the BIC is partially populated, and rest of the pump excitations eventually decay into the signal continuum, following the arguments made in Sec.~\ref{sec:singlephotonPDC}.

\section{PDC beyond the weak-excitation regime} \label{sec:general}
\begin{figure*}[ht]
    \centering
    \includegraphics[width=0.99\textwidth]{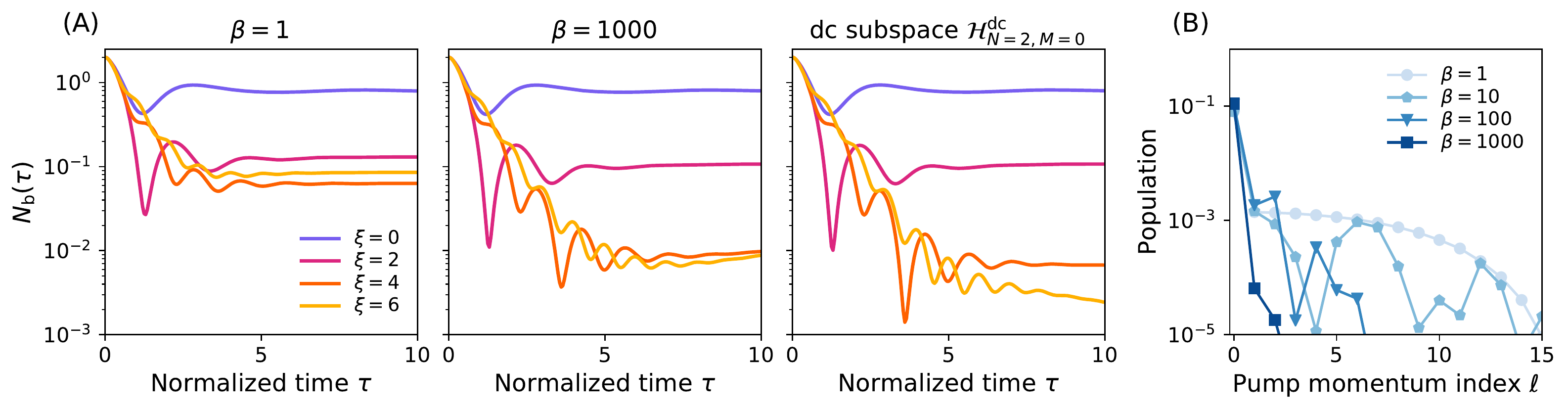}
    \caption{(A) Numerically simulated evolution of the pump photon population $N_\text{b}(\tau)$ for an input two-photon dc pump state $\ket{N=2} = \frac{1}{\sqrt2}\hat b_0^{\dagger2}\ket{0}$ undergoing PDC at different normalized detuning (i.e., phase mismatch) $\xi$ and relative second-order dispersion (i.e., group velocity dispersion) $\beta = d_\text{b}/d_\text{a}$. In both the left ($\beta=1$) and the middle ($\beta=1000$) figures, the normalized first-order dispersion (i.e., group velocity mismatch) $\gamma = 0$; in the right figure, the simulation is restricted to the dc pump subspace $\mathcal{H}^\text{dc}_{N=2,M=0}$ (see \eqref{eq:dc_subspace}), corresponding to the case of $\beta\rightarrow\infty$ or $\gamma\rightarrow\infty$. (B) Population of pump states $\hat{b}_{\ell}^\dagger\hat{b}_{-\ell}\ket{0}$ at normalized time $\tau = 10$ for various $\beta$ and $\xi=\gamma=0$, showing that non-dc pump excitations ($\ell > 0$) decrease as the dispersion $\beta$ increases, indicating confinement of dynamics to $\ell = 0$. For all simulations, $\epsilon=1/30$ is used, corresponding to quantization window length $L=30\zeta$.}
    \label{fig:evolution_2p}
\end{figure*}
When the excitation of the input pump mode is not small enough to assume the regime of weak excitation, we need to take into account interactions between multiple pump photons and the multiple downconverted signal pairs. In this section, we consider the dynamics induced by such multi-photon interactions. Extending the discussion from the beginning of Sec.~\ref{sec:singlephotonPDC}, an input monochromatic coherent state with photon flux density $\rho$ (taken to be dc or $\ell = 0$ for simplicity) takes the form $\exp\bigl(\sqrt{\rho L}(\hat{b}^\dagger_0-\hat{b}_0)\bigr)\propto\sum_N\sqrt{(\rho L)^N/N!}\ket{N}$, where $\ket{N}=\frac{1}{\sqrt{N!}}\hat{b}_0^{\dagger N}\ket{0}$ is an $N$-photon (dc) pump state. (As usual, $L$ is the length of the quantization window, chosen to fully contain spatial features with characteristic length scale $L_\text{c}$.) When $\rho L_\text{c}\not\ll 1$, $\rho L$ is not necessarily small (because of $L>L_\text{c})$, so contributions from multi-photon states $\ket{N\geq2}$ become non-negligible. In this beyond-weak-excitation regime, multi-photon processes can generally populate pump states with a finite spread in momentum despite the initial state being monochromatic, and we can no longer assume the pump mode is well-described by considering only a single discrete state. Nevertheless, as we show in this section, viewing these multi-photon effects as Fano-type discrete-continuum interactions still provides useful intuition for understanding the PDC dynamics under certain conditions.

For multi-photon dynamics, the dimension of the Hilbert space we need to consider grows exponentially with $N$. For an initial $N$-photon pump state, the relevant states reached by its evolution are those of the form
\begin{align}
\label{eq:basis}
\begin{split}
\ket{\mathcal{M}_{2N-2J},\mathcal{L}_J}\propto\hat{a}_{m_1}^\dagger\dots\hat{a}_{m_{2N-2J}}^\dagger \hat{b}_{\ell_1}^\dagger\dots \hat{b}_{\ell_J}^\dagger\ket{0},
\end{split}
\end{align}
where the pump population $J \in \{0,1,\dots,N\}$, and $\mathcal{M}_{2N-2J}=\{m_1,m_2,\dots,m_{2N-2J}\}$ ($m_\alpha \in \mathbb Z$) and $\mathcal{L}_J=\{\ell_1,\ell_2,\dots,\ell_{J}\}$ ($\ell_\alpha \in \mathbb Z$) are multiset labels. Such states comprise all the eigenstates of $\hat{N}$ with eigenvalue $N$, and we denote the subspace of the Hilbert space spanned by these states as $\mathcal{H}_N$, whose dimension scales as $\sim m_\text{max}^{2N}$ where $m_\text{max}$ is a chosen cutoff for the momentum index. Because the total momentum $\hat{M}$ is another conserved quantity, we can independently consider the dynamics within each subspace $\mathcal{H}_{N,M}$ of $\mathcal{H}_N$ spanned by eigenstates of $\hat{M}$ with eigenvalues $M$. More specifically, $\mathcal{H}_{N,M}$ is spanned by all the states of the form \eqref{eq:basis} that fulfill the condition $\sum_{i=1}^{2N-2J}m_i+\sum_{j=1}^J \ell_j=M$, and the dimension of $\mathcal H_{N,M}$ therefore scales as $\sim m_\text{max}^{2N-1}$.

To obtain numerical models for the dynamics, we compute the matrix elements of the Hamiltonian \eqref{eq:fullhamiltonian} on states in the subspace $\mathcal{H}_{N,M}$. The diagonal elements are
\begin{subequations}
\begin{align}
\begin{split}
    &\bra{\mathcal{M}_{2N-2\nu},\mathcal{L}_\nu}\hat{H}\ket{\mathcal{M}_{2N-2\nu},\mathcal{L}_\nu}/\kappa\hbar\\
    &=\sum_{i=1}^{2N-2\nu}\frac{\epsilon^2}{2} m_i^2+\sum_{j=1}^J \left(\xi+\gamma\epsilon \ell_j+\frac{\beta\epsilon^2}{2}\ell_j^2\right)
\end{split}
\end{align}
where $\gamma=\mu/(2 d_\text{a}g^2)^\frac{1}{3}$ and $\beta=d_\text{b}/d_\text{a}$ are both independent of $L$. The nonzero off-diagonal elements are
\begin{align}
\label{eq:offdiag_n}
\begin{split}
    &\bra{\mathcal{M}'_{2N-2J'},\mathcal{L'}_{J'}}\hat{H}\ket{\mathcal{M}_{2N-2J},\mathcal{L}_{J}}/\kappa\hbar\propto\epsilon^\frac{1}{2},
\end{split}
\end{align}
\end{subequations}
which exist only when $J=J'+1$ or $J'=J+1$ and $\mathcal{M}_{2N-2J}\subset \mathcal{M}'_{2N-2J'}$ or $\mathcal{M}'_{2N-2J'}\subset \mathcal{M}_{2N-2J}$, respectively. Here, the proportional symbol takes into account the multiplicity of the interaction; we provide explicit expressions for this factor in Appendix~\ref{sec:multiplicity}.

In Fig.~\ref{fig:evolution_2p}(A), we show the result of numerical simulations of the evolution of an initial two-photon dc pump state $\ket{N=2} = \frac1{\sqrt2} \hat b_0^{\dagger2}\ket{0}$ within the subspace $\mathcal{H}_{N=2,M=0}$. In contrast to the single-photon (i.e., weak-excitation) case, the terminal pump photon population $\lim_{\tau\rightarrow\infty}N_\text{b}(\tau)$ does not decrease monotonically with the detuning $\xi$, suggesting that multi-photon interactions can suppress the full pump depletion expected for the single-photon case in the dissipative regime $\xi \rightarrow \infty$. It is worth mentioning that, beyond the weak excitation regime, pump excitations are not necessarily confined to the initial dc mode as in the single-photon case. For instance, a sequence of allowed transitions $\hat{b}_0^{\dagger2}\ket{0}\rightarrow\hat{a}_{m_1}^\dagger\hat{a}_{-m_1}^\dagger\hat{b}_0^\dagger\ket{0}\rightarrow\hat{a}_{m_1}^\dagger\hat{a}_{-m_1}^\dagger\hat{a}_{m_2}^\dagger\hat{a}_{-m_2}^\dagger\ket{0}\rightarrow \hat{a}_{m_1}^\dagger\hat{a}_{m_2}^\dagger\hat{b}_{-m_1-m_2}^\dagger\ket{0}\rightarrow\hat{b}_{m_1+m_2}^\dagger\hat{b}_{-m_1-m_2}^\dagger\ket{0}$ lead to a spreading in the momentum of excited pump modes. As a result, the pump degree of freedom can no longer be seen as discrete, and thus, the intuitive picture of the PDC process as a purely Fano-type discrete-continuum interaction no longer applies.

It is worth noting, however, that such momentum spreading during backconversion can be suppressed in the limit of large pump dispersion, i.e., $\gamma\rightarrow\infty$ or $\beta\rightarrow\infty$. For example, for the initial two-photon dc pump state $\ket{N=2}$, non-dc pump states (i.e., with $\ell \neq 0$) remain unpopulated throughout the evolution since they are energetically prohibited (i.e., they have large phase-mismatch). This can be seen in Fig.~\ref{fig:evolution_2p}(B) showing the population of pump photons $\hat{b}^\dagger_{\ell}\hat{b}^\dagger_{-\ell}\ket{0}$ for various $\beta$, in which we observe the population of the non-dc pump modes ($\ell\geq1$) are suppressed when $\beta$ becomes large. Formally we expect dynamics in this limit are contained within a ``dc pump subspace'' $\mathcal{H}^\text{dc}_{N,M=0}$ spanned by
\begin{align}
\label{eq:dc_subspace}
    \ket{\widetilde{\mathcal{M}}_{N-J}}\propto\Biggl( \prod_{j=1}^{N-J}\hat{a}^\dagger_{m_j}\hat{a}^\dagger_{-m_j}\Biggr)\hat{b}_0^{\dagger J}\ket{0}
\end{align}
where $\widetilde{\mathcal{M}}_{N-J}=\{m_1,\dots,m_{N-J}\}$ is a multiset label containing only the momentum indices of signal photons. (Note that we label this subspace as dc since our initial state occupies $\ell = 0$, but in principle the same construction can be made for an initial state with arbitrary $\ell$ in the limit of high pump dispersion.)

In Fig.~\ref{fig:evolution_2p}(A), we show the time evolution of the pump photon population $N_\text{b}(\tau)$ for the initial state $\ket{N=2}$ for various values of $\beta$, with numerical simulation done in both the full subspace $\mathcal H_{N=2,M=0}$ as well as using only the dc subspace $\mathcal{H}^\text{dc}_{N=2,M=0}$. Notably, the pump depletion rate is much higher and decreases monotonically with the detuning $\xi$ when the pump excitations are confined to the dc mode, as expected from the arguments made in Sec.~\ref{sec:singlephotonPDC} for dissipative dynamics based on Fano-type discrete-continuum interactions. The numerical results also show that, as argued above, the dynamics simulated within the dc subspace become closer to that simulated in the full subspace as $\beta$ increases.

\begin{figure}[tb]
\begin{center}
    \includegraphics[width=0.43\textwidth]{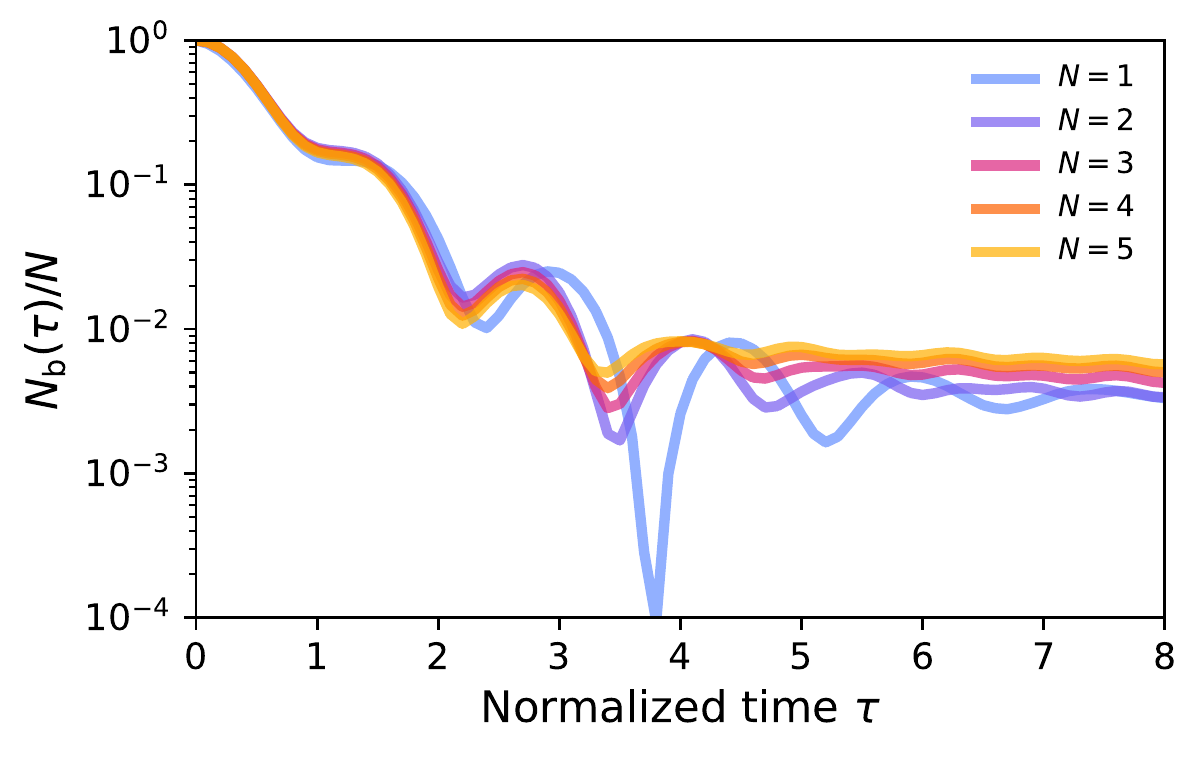}
    \caption{Numerically simulated pump population dynamics for multi-photon input dc pump states $\ket{N} \propto \hat b_0^{\dagger N}\ket{0}$, assuming confinement of the dynamics to the dc subspace $\mathcal{H}^\text{dc}_{N,M=0}$ (see \eqref{eq:dc_subspace}, valid in the limit of large pump dispersion). For these simulations, we use $\epsilon=1/20$ and normalized detuning $\xi=4$.}
    \label{fig:evolution_np}
    \end{center}
\end{figure}

An interesting consequence is that the complexity of the numerical simulation reduces considerably in the limit of large pump dispersion, because the dimension of $\mathcal{H}^\text{dc}_{N,M=0}$ only scales as $\sim m_\text{max}^{N}$ as opposed to $\sim m_\text{max}^{2N-1}$ for the full subspace $\mathcal{H}_{N,M=0}$. Thus, the reduced model using the dc subspace allows us to perform numerical simulations for this particular limit with a larger photon number $N$. In Fig.~\ref{fig:evolution_np}, we show the time evolution of the pump photon population for $\ket{N}$ up to $N=5$, where we observe that the pump is asymptotically depleted to less than \SI{1}{\percent}, again in accordance with our intuition from Sec.~\ref{sec:singlephotonPDC} for dissipative PDC based on Fano's theory.

\section{Extensions to higher-order parametric interactions} \label{sec:chi3}
While we have largely focused on PDC in 1D $\chi^{(2)}$ waveguides in this work, this framework for analyzing photon downconversion using Fano's theory of discrete-continuum interactions can be extended to more general scenarios, including higher-order spatial dimensions and higher-order nonlinear interactions. In this section, as a natural extension of our theoretical framework, we study a quantum model for single-photon pumped three-photon generation (TPG), where a third-harmonic pump photon downconverts to a triplet of fundamental-harmonic signal photons in the weak excitation regime. Using Fano's theory, we again obtain a complete solution for the TPG Hamiltonian eigenproblem including both bound and continuum eigenstates. We show that, as for the case of $\chi^{(2)}$ PDC, the dynamics of the pump photon population for TPG exhibit characteristic dissipative/dispersive dynamics depending on the nature of the coupling.

Here, as a model for broadband parametric three-photon generation, we consider a (rotating frame; see Sec.~\ref{sec:model}) Hamiltonian of the form
\begin{subequations} \label{eq:chi3hamiltonian}
\begin{small}\begin{align}
\label{eq:nl3tot}
\frac{\hat{H}_{\text{NL}}}{\hbar}+\sum^\infty_{m=-\infty} d_\text{a}m^2\hat{a}_m^\dagger\hat{a}_m+\sum^\infty_{\ell=-\infty}(\delta+\mu\ell+d_\text{c}\ell^2)\hat{c}_\ell^\dagger\hat{c}_\ell,
\end{align}\end{small}with nonlinear part
\begin{align}
\label{eq:nl3}
\hat{H}_{\text{NL}}/\hbar=\frac{\chi}{3}\sum_{m_1+m_2+m_3=\ell}\left(\hat{a}_{m_1}^\dagger\hat{a}_{m_2}^\dagger\hat{a}_{m_3}^\dagger\hat{c}_{\ell}+\mathrm{H.c.}\right),
\end{align}
\end{subequations}
where $\hat{a}_m$ and $\hat{c}_\ell$ are the annihilation operators for the signal and pump modes with momentum indices $m$ and $\ell$, respectively; $\delta$ is the phase-mismatch; $\mu\propto L^{-1}$ represents the group velocity mismatch;  and $d_\text{a}, d_\text{c}\propto L^{-2}$ represent the group velocity dispersion of the signal and pump respectively. The scaling of the dispersion and nonlinear coupling rates with the quantization window length $L$ are derived as in Sec.~\ref{sec:model} for PDC (e.g., \eqref{eq:dispersions} for the dispersion); here the nonlinear coupling rate $\chi \propto L^{-1}$.

Note that the model \eqref{eq:chi3hamiltonian} does not include the self-phase modulation terms $\propto\sum_{m_1+m_2=m_3+m_4}\hat{a}_{m_1}^\dagger\hat{a}_{m_2}^\dagger\hat{a}_{m_3}\hat{a}_{m_4}$, which generally coexist with $\hat{H}_{\text{NL}}$ when TPG is supported by a native material $\chi^{(3)}$. While we only focus on the parametric part of the interaction here, an accurate model for a realistic system with native $\chi^{(3)}$ may require a more rigorous treatment of the self-phase modulation, which is beyond the scope of this work. The idealized model \eqref{eq:chi3hamiltonian} can be appropriate, for instance, in the case where self-phase modulation is compensated with cascaded $\chi^{(2)}$ nonlinear processes~\cite{Beckwitt2001, Dorrer2014, Kobyakov1996}.

As another subtlety in the model \eqref{eq:chi3hamiltonian}, special care should also be taken due to the presence of a singularity in which the dynamics can feature infinitely broadband signal eigenstates, resulting in point-like spatial wavefunctions in the dynamics. This singularity is the result of ignoring higher-order dispersion, and it also arises in $\chi^{(2)}$ models with more than one spatial dimension and no higher-order dispersion~\cite{Kheruntsyan2000, Kheruntsyan1998, Kheruntsyan1998b}. In Ref.~\cite{Kheruntsyan2000, Kheruntsyan1998, Kheruntsyan1998b}, it has been shown that one way to approach this issue is to formally impose a finite momentum cutoff, which produces an effective model that eliminates the singularity in a manner similar to that of higher-order dispersion. In this work, we employ the same treatment, where we only keep terms with $K=\frac{2\pi}{L}\sqrt{m_1^2+m_2^2+m_3^2}\leq K_\text{max}$ in the nonlinear term \eqref{eq:nl3}.

Following the same procedure as in Sec.~\ref{sec:singlephotonPDC}, we suppose the initial state is a single-photon pump state $\ket{c_0}=\hat{c}^\dagger_0\ket{0}$ (in the dc mode $\ell = 0$ without loss of generality). To obtain the dynamics, we focus on the matrix elements of \eqref{eq:chi3hamiltonian} within the subspace spanned by states accessible from $\ket{c_0}$ via TPG and then take a continuum limit to obtain analytic expressions for the eigenenergies and eigenstates of the system. As the derivation of the continuum Hamiltonian is rather involved, we present the full details in Appendix~\ref{sec:threephoton}. The result of that calculation is that the continuum-limit single-photon-TPG Hamiltonian can be written in the form $\hbar\kappa\hat G^\text{cont}$, where the effective nonlinear coupling rate (cf. \eqref{eq:kappa} for PDC) is
\begin{align}
    \kappa=\frac{2\chi^2}{d_\text{a}},
\end{align}
and the normalized Hamiltonian takes the form
\begin{align}
    \label{eq:gcont_maintext}
        &\hat{G}^\text{cont}=\xi\hat{w}^\dagger\hat{w}\\
    &+\frac{2}{\sqrt{3}}\int^{r_\text{max}}_0\mathrm{d}r\left[r^2\hat{\phi}^\dagger_{r}\hat{\phi}_{r}+\sqrt{\frac{\pi r}{3}}\left(\hat{\phi}_{r}\hat{w}^\dagger+\hat{\phi}_{r}^\dagger\hat{w}\right)\right].\nonumber
\end{align}
Here $\hat{w}=\ket{0}\bra{c_0}$ annihilates a dc pump photon with normalized detuning $\xi=\delta/\kappa$, while $\hat{\phi}_r$ (see \eqref{eq:superposition} in Appendix~\ref{sec:threephoton}), with commutation relation $[\hat{\phi}_r,\hat{\phi}_{r'}^\dagger]=\delta(r-r')$, annihilates an excitation composed of a uniform superposition of photon-triplet states with normalized momentum $r=K\zeta/2\sqrt{2}\pi$, where $\zeta=d_\text{a}L/\chi$ is the characteristic correlation length (independent of $L$). As discussed above, a momentum cutoff is imposed via $r_\text{max}=K_\text{max}\zeta/2\sqrt{2}\pi$.

Again following Ref.~\cite{Fano1961}, we posit that the eigenstates of $\hat{G}^\text{cont}$ with eigenvalue $\lambda$ take the form
\begin{align}
\ket{\varphi_{\lambda}}=\left(c_{\lambda}\hat{w}^\dagger+\int^{r_\text{max}}_0\mathrm{d}rf_{\lambda}(r)\hat{\phi}_r^\dagger\right)\ket{0},
\end{align}
which leads to
\begin{subequations}
    \begin{align}
        \label{eq:chi3c}
        \xi c_{\lambda}+\int^{r_\text{max}}_0\mathrm{d}r\frac{2\sqrt{\pi r}}{3}f_{\lambda}(r)&=\lambda c_{\lambda}\\
        \label{eq:chi3f}
        \frac{2}{\sqrt{3}}r^2f_{\lambda}(r)+\frac{2\sqrt{\pi r}}{3}c_{\lambda}&=\lambda f_{\lambda}(r).
    \end{align}
\end{subequations}

For negative eigenvalues $\lambda=-\lambda_{\text{T}}<0$, where the subscript ``T'' denotes ``three-photon bound state'', we obtain a single (i.e., bound) solution to \eqref{eq:chi3f}
\begin{align}
    \label{eq:fc}
    f_{\text{T}}(r)=-\frac{2\sqrt{\pi r}c_{\text{T}}}{2\sqrt{3}r^2+3\lambda_{\text{T}}},
\end{align}
where $c_{\text{T}} = c_{-\lambda_\text{T}}$ and $f_{\text{T}} = f_{-\lambda_\text{T}}$. Substitution of \eqref{eq:fc} into \eqref{eq:chi3c} yields an equation that determines the energy of the three-photon bound state:
\begin{align}
    \frac{\pi}{3\sqrt{3}}\log \left[1+\frac{2r_\text{max}^2}{\sqrt{3}\lambda_{\text{T}}}\right]-\lambda_{\text{T}}=\xi.
\end{align}
Finally, normalization of the bound state gives
\begin{align}
    c_{\text{T}}^2=\left(1+\frac{2\pi r_\text{max}^2}{3\lambda_{\text{T}}\left(2\sqrt{3}r_\text{max}^2+3\lambda_{\text{T}}\right)}\right)^{-1},
\end{align}
which completes the bound state solution.

For positive energies $\lambda>0$, we have from \eqref{eq:chi3f}
\begin{small}\begin{align}
    \begin{split}
    f_{\lambda}(r)=\frac{2\sqrt{\pi r}c_{\lambda}}{3}&\left[\frac{1}{\lambda-\frac{2r^2}{\sqrt{3}}}+w(\lambda)\delta\left(\lambda-\frac{2r^2}{\sqrt{3}}\right)\right]
    \end{split},
\end{align}\end{small}where $w(\lambda)$ is to be determined so that \eqref{eq:chi3c} is fulfilled, which yields
\begin{align}
    w(\lambda)=\frac{3\sqrt{3}}{\pi}(\lambda-\xi)+\log\left[\frac{2r_\text{max}^2}{\sqrt{3}\lambda}-1\right].
\end{align}
Enforcing the normalization $\langle \varphi_{\lambda}\vert \varphi_{\lambda'}\rangle=\delta(\lambda-\lambda')$ gives
\begin{align}
    c_{\lambda}^2=\frac{3\sqrt{3}}{\pi}\frac{1}{\pi^2+w^2(\lambda)},
\end{align}
which completes the continuum solution.

\begin{figure}[bth]
    \includegraphics[width=0.45\textwidth]{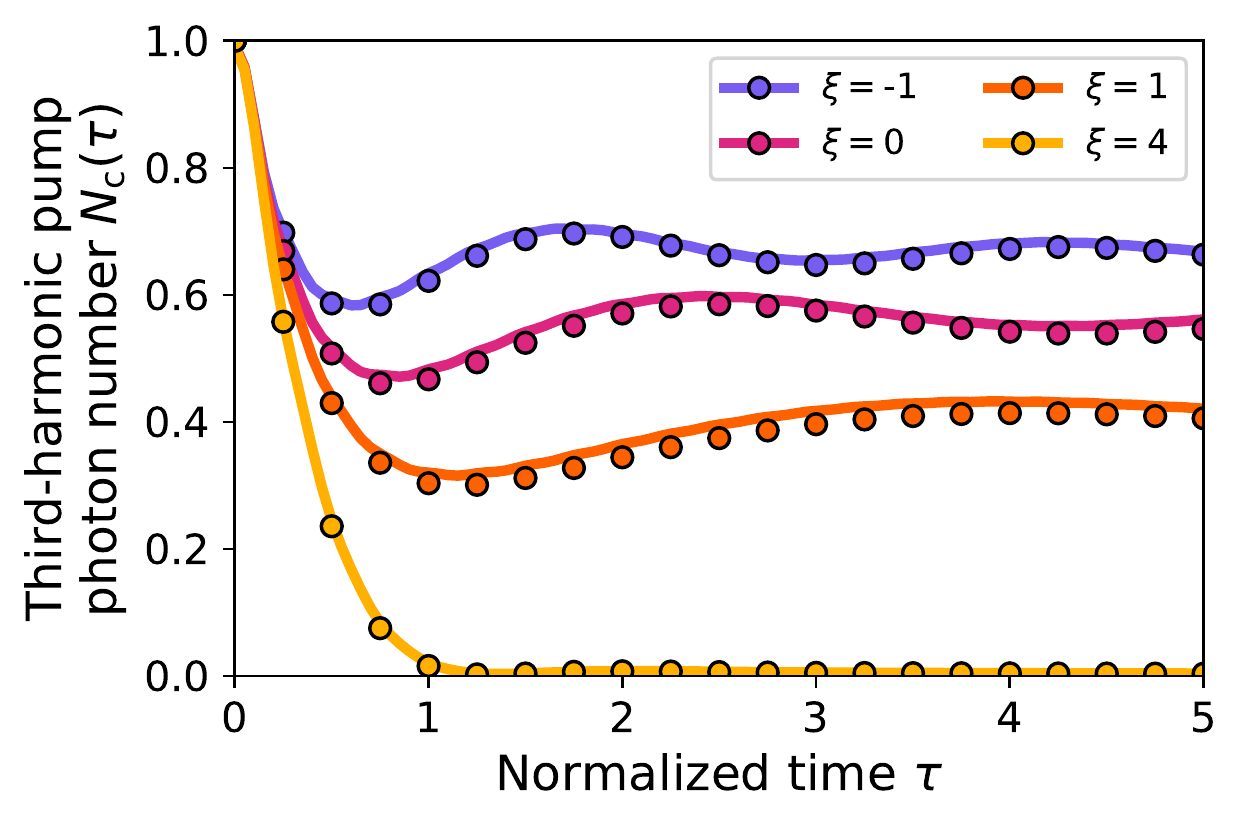}
    \caption{Third-harmonic pump photon population $N_{\text{c}}(\tau)$ as a function of (normalized) propagation time $\tau$ for various (normalized) detunings $\xi$ (i.e., phase mismatch), ranging from dispersive ($\xi < 0$) to dissipative ($\xi > 0$). Solid lines are based on analytic evaluation of \eqref{eq:nevlovechi}, while circles show numerical results based on simulating \eqref{eq:chi3hamiltonian} with finite quantization window length ($\epsilon = 1/100$). As discussed in the main text, a momentum cutoff $r_\text{max}=5$ is imposed in this system.}
    \label{fig:fourwave}
\end{figure}

Combining these results (cf.\ \eqref{eq:evolution} for PDC), the pump population as a function of normalized time $\tau=\kappa t$ is
\begin{align}
\label{eq:nevlovechi}
N_{\text{c}}(\tau)=\left| c_{\text{T}}^2+\int_0^{r_\text{max}^2}\frac{3\sqrt{3}e^{-\mathrm{i}(\lambda+\lambda_{\text{T}})\tau}}{\pi\left(\pi^2+w^2(\lambda)\right)}\,\mathrm{d}\lambda\right|^2.
\end{align}
In Fig.~\ref{fig:fourwave}, we show $N_{\text{c}}(\tau)$, both according to the analytic result \eqref{eq:nevlovechi} corresponding to $L\rightarrow \infty$ as well as according to numerical simulation of the discrete Hamiltonian \eqref{eq:chi3hamiltonian} with finite $L$, obtaining good agreement for $\epsilon \ll 1$. As was the case for PDC, the qualitative nature of the TPG interaction changes from dispersive to dissipative as the detuning $\xi$ varies from negative to positive, resulting in damped-Rabi-like energy exchange for the former case and monotonic decay of the third-harmonic pump population for the latter.
\section{Prospects for Experiments} \label{sec:experiments}
In this section, we relate the parameters of our quantum model to experimentally relevant parameters to better understand the physical regimes in which such Fano-type quantum dynamics can occur. We focus on 1D $\chi^{(2)}$ waveguides and relate the nonlinear rate $g$ to the second-harmonic generation (SHG) slope conversion efficiency $\eta$ widely used as a classical figure of merit for such devices, resulting in a quantum figure of merit which can be evaluated in emerging platforms for quantum nanophotonics.

We consider a 1D $\chi^{(2)}$ nonlinear waveguide in which the phase and group velocity matching conditions are met, i.e., $\delta=\mu=0$ (see \eqref{eq:dispersions} for dispersion parameters). Here, the mean field equations of motion induced by the Hamiltonian \eqref{eq:fullhamiltonian} with a quantization window length $L$ are
\begin{subequations}
\label{eq:shgeom}
\begin{align}
    \frac{\mathrm{d}\alpha_m}{\mathrm{d}t}&=-\mathrm{i}g\sum_{\ell=n+m}\alpha_n^*\beta_\ell-\mathrm{i}d_\text{a}m^2\alpha_m\\
     \frac{\mathrm{d}\beta_\ell}{\mathrm{d}t}&=-\frac{\mathrm{i}g}{2}\sum_{\ell=n+m}\alpha_m\alpha_{n}-\mathrm{i}d_\text{b}\ell^2\beta_\ell,
\end{align}
\end{subequations}
where we have formally made the usual mean-field-limit c-number substitutions $\hat{a}_m\mapsto \alpha_m$ and $\hat{b}_\ell\mapsto \beta_\ell$. We interpret these c-numbers as indicating that the optical power in each wavevector bin centered on $k_{a0}+2\pi m/L$ and $k_{b0}+2\pi \ell/L$ (with wavevector bandwidths $2\pi/L$) is given by $P_{\text{a}m}=\hbar \omega(k_{\text{a}m})|\alpha_m|^2v/L$ and $P_{\text{b}\ell}=\hbar \omega(k_{\text{b}\ell})|\beta_\ell|^2v/L$ for the signal and pump, respectively. Here, $v = \omega'(k_{\text{a}0})$ denotes the group velocity at the signal carrier, and we recall $k_{\text{b}0} = 2k_{\text{a}0}$ as defined in Sec.~\ref{sec:model}.

We consider an initial classical state $\alpha_m=\alpha_0(0)\delta_{m,0}$ and $\beta_\ell=0$ for all $\ell$, which results in SHG dynamics starting from a monochromatic coherent state at the carrier (i.e., dc mode) of the fundamental-harmonic band. As can be seen from \eqref{eq:shgeom}, only $\alpha_0$ and $\beta_0$ undergo nontrivial time evolution. Thus, the resulting dynamics are captured by the ``cw'' classical SHG equations of motion
\begin{align}
\label{eq:reducedshgeom}
    &\frac{\mathrm{d}\alpha_0}{\mathrm{d}t}=-\mathrm{i}g\alpha_0^*\beta_0,
     &\frac{\mathrm{d}\beta_0}{\mathrm{d}t}=-\frac{\mathrm{i}g}{2}\alpha_0^2,
\end{align}
whose solutions~\cite{Armstrong1962} are
\begin{subequations}
\begin{align}
\alpha_0(t)&=\alpha_0(0)\sech\left(\frac{\alpha_0(0)gt}{\sqrt{2}}\right)\\
\beta_0(t)&=-\frac{\mathrm{i}\alpha_0(0)}{\sqrt{2}}\tanh\left(\frac{\alpha_0(0)gt}{\sqrt{2}}\right).
\end{align}
\end{subequations}
The SHG slope conversion efficiency can be defined as
\begin{align}
\eta=\frac{1}{2v^2}\left.\frac{\mathrm{d}^2}{\mathrm{d}t^2}\frac{P_{\text{b}0}(t)}{P_{\text{a}0}^2(0)}\right|_{t=0}=\frac{g^2L}{2\hbar\omega(k_{\text{a}0})v^3},
\end{align}
which establishes the relationship between $g$ and $\eta$ for a given $L$. To interpret this result independently of $L$, we recall the characteristic nonlinear rate $\kappa$ from \eqref{eq:kappa} in order to define a characteristic propagation length
\begin{align}
\label{eq:lpdc}
    L_\text{PDC}=\frac{v}{\kappa} = \left(\frac{\lambda^2|k_\text{a}''|}{4\hbar^2c^2\eta^2}\right)^\frac{1}{3},
\end{align}
where $\lambda = 2\pi c/\omega_{\text{a}0}$ is the wavelength of the signal carrier. Here, $d_\text{a}$ occurring in the definition of $\kappa$ has been related to the group velocity dispersion $k_\text{a}''=k_{\text{a}0}''\bigl(\omega(k_{\text{a}0})\bigr)$ of the signal carrier using $d_\text{a} = -2\pi^2v^3k_\text{a}''/L^2$. We take the absolute value of $k_\text{a}''$ as the quantum dynamics of \eqref{eq:fullhamiltonian} are insensitive to the sign of $d_\text{a}$, after appropriate remapping of parameters (see Sec.~\ref{sec:singlephotonPDC}). Note that for a given material system, scaling the waveguide dimensions with the wavelength results in $A_\text{eff}\sim \lambda^{-2}$~\cite{Jankowski2020} and thus $\eta\sim\lambda^{-4}$, from which it follows that $L_\text{PDC}\sim \lambda^{10/3}$; this strong scaling with wavelength underscores the importance of pushing dispersion engineering into the regime of shorter wavelengths. It is also worth mentioning that \eqref{eq:lpdc} assumes the system bandwidth is limited by signal group velocity dispersion $k_\text{a}''$ as our model only includes dispersion up to second order (see Sec.~\ref{sec:model}); similar expressions for $L_\text{PDC}$ (possibly with different scaling) can be obtained using quantum models incorporating higher-order dispersion. Advancements in dispersion engineering to flatten energy dispersion over an extended bandwidth may allow for improvements in $L_\text{PDC}$ beyond \eqref{eq:lpdc}. 

The length $L_\text{PDC}$ is related (via $v$) to the characteristic time of propagation $1/\kappa$ required for single-photon PDC to exhibit the dynamics analyzed in Sec.~\ref{sec:singlephotonPDC} (and later). Thus, as an experimental figure of merit, $L_\text{PDC}$ indicates the length-scale requirements for physical 1D $\chi^{(2)}$ nonlinear waveguides to exhibit such dynamics. For state-of-the-art nanophotonic waveguides in thin-film periodically-poled $\text{LiNbO}_3$ (PPLN)~\cite{Luo2018,Zhang2019,Wang2018,Jankowski2020} operating at a fundamental wavelength $\lambda=\SI{1.5}{\micro\meter}$, Ref.~\cite{Wang2018} has demonstrated a slope conversion efficiency of $\eta=\SI{2600}{\percent\per\watt\per\centi\meter\squared}$, while on GaAs platforms~\cite{Kuo2014, Stanton2020}, Ref.~\cite{Stanton2020} has demonstrated $\eta=\SI{47000}{\percent\per\watt\per\centi\meter\squared}$ at $\lambda=\SI{2.0}{\micro m}$ for GaAs-on-insulator. Assuming realistic engineered group velocity dispersions $k_\text{a}''\sim\SI{-5}{\femto\second\squared\per\milli\meter}$, we obtain respective characteristic lengths of $L_\mathrm{PDC}\sim\SI{3.5}{m}$ and $L_\mathrm{PDC}\sim\SI{60}{cm}$. 

While these length scales may still be geometrically unrealistic for current-generation nanophotonic devices with straight waveguides on mm-scale chips, it is worth noting that, from a decoherence perspective, current figures of merit are now approaching a critical breakeven point where $L_\text{PDC}$ and the characteristic decoherence length are commensurate, at least for the thin-film PPLN platform where a \num{3}-\si{dB} attenuation length as long as \SI{1}{m}~\cite{Zhang2017} (i.e., \SI{3}{\decibel \per\meter} power attenuation) has already been demonstrated (vs.\ $L_\text{PDC} \sim \SI{3.5}{m}$). The remaining
challenge thus seems to lie either in the development of more sophisticated fabrication techniques to overcome geometric constraints for current chip architectures---such as being able to etch low-loss curved nonlinear waveguides of adequate geometrical uniformity~\cite{Lee2012} or introduce resonant structures~\cite{Zhang2017} to recycle the physical interaction length---or in advanced dispersion engineering to shorten the operating wavelength (potentially to just below the material bandgap) and enhance the nonlinearity accessible. In this context, the potential to see unique broadband dynamics emerge in physics as conceptually foundational as parametric downconversion may serve as a convenient target to guide photonic engineering, fabrication, and modeling efforts as they finally begin to tackle head-on the experimental challenge of accessing the quantum regime of broadband nonlinear nanophotonics.

\section{Conclusions} \label{sec:conclusions}
We have shown that Fano's theory for discrete-continuum interactions can provide a unified theoretical framework for analyzing broadband PDC of a weak pump field in the highly nonlinear regime of a 1D $\chi^{(2)}$ waveguide. Within this theoretical framework, we have derived analytic results characterizing, both qualitatively and quantitatively, the quantum dynamics of few-photon broadband PDC while providing physical analogies to other multimode systems known to exhibit Fano-type physics, such as atomic/molecular autoionization. An extended analysis of two coupled nonlinear waveguides has revealed even richer physics arising from Fano-type interactions, including the suppression/enhancement of PDC due to destructive/constructive interference between multiple discrete-continuum couplings and the emergence of a BIC under appropriate conditions. Beyond $\chi^{(2)}$-based single-photon PDC, three-photon generation can also be treated by our theory, and we have seen numerical evidence of Fano-like dissipative dynamics even after including multi-photon interactions under stronger pumping. While our results have immediate implications for guiding experimental research in the development of next-generation nanophotonic devices, the broader perspectives offered by Fano's theory may also help address the inherent challenges associated with the theoretical and model intractability of strongly nonlinear broadband quantum optical systems, opening a pathway towards exploiting such systems as a platform for quantum engineering.

\acknowledgments
This work has been supported by the Army Research Office under Grant No.\ W911NF-16-1-0086, the National Science Foundation under awards CCF-1918549 and PHY-2011363, and SLAC program SLAC DOE/SU Contract DE-AC02-76SF00515. The authors also wish to thank NTT (Nippon Telegraph and Telephone Corporation) Research for their financial and technical support. R.\,Y.\ is also supported by Masason foundation. R.\,Y.\ would like to thank Tomohiro Soejima and Atsushi Yamamura for helpful discussions.
\begin{appendix}

\section{ESSENTIAL PHYSICS OF DISCRETE-CONTINUUM INTERACTIONS}
In this section, we provide some general remarks that summarize, on a conceptual/schematic level, the approach we take in analyzing single-photon PDC as a discrete-continuum interaction, essentially by recasting the physics into the form first considered by Ugo Fano in studying atomic autoionization~\cite{Fano1961}.

Generically, a system featuring discrete-continuum interactions can be characterized by a (normalized) Hamiltonian $\hat G$ which couples together a set of ``bare'' discrete states $\ket{\phi_i}$, where $i$ is a discrete label, with a set of bare continuum states $\ket{\phi_x}$, where $x$ is a continuous label. Fano's theory can then be applied to solve for the system eigenstates $\ket{\varphi_\lambda}$ with eigenenergy $\lambda$, which generically take a ``dressed'' or ``hybridized'' form
\begin{equation} \label{eq:generic-dressed}
\ket{\varphi_\lambda} = \sum_i {c_{i,\lambda}} \ket{\phi_i} + \int \mathrm{d}x \, f_\lambda(x) \ket{\phi_x}.
\end{equation}
Typically, one finds after this analysis a set of discrete eigenstates $\ket{\varphi_{\lambda_k}}$ together with a set of continuum eigenstates $\ket{\varphi_\lambda}$. For example, there may be a critical $\lambda_\text{c}$ so that for eigeneneriges less than $\lambda_\text{c}$, there exist only discrete eigenstates, while for eigenenergies greater than $\lambda_\text{c}$, there is a ``band'' of continuum eigenstates.

When the system parameters are tuned such that, for example, the gap between the highest-energy discrete eigenstate and the lowest-energy continuum eigenstate is large, the dressed states tend to feature less hybridization among the bare states of the system, and an initial excitation along one of the bare discrete states, say $\ket{\phi_0}$ may only have significant overlap with one of the discrete eigenstates, say $\ket{{\varphi_\lambda}_0}$, in which case the continuum eigenstates play a negligible role in the evolution, and the excitation remains bound. In analogy with the dispersive-coupling limit of cavity quantum electrodynamics where the dressed states feature weak hybridization of the bare states, we call this the ``dispersive'' limit.

On the other hand, in autoionization, dissipative single-photon PDC, or indeed many other Fano-type processes, a system initialized in one of its discrete bare states, say $\ket{\phi_0}$, may have significant overlap with the continuum eigenspace $\Span\{\ket{\varphi_\lambda} : \lambda > \lambda_\text{c}\}$. This overlap can be characterized by introducing a quantity
\begin{align}
\label{eq:bigflambda}
|F_{\lambda}|^2=|\langle\varphi_\lambda\vert\phi_0\rangle|^2=|c_{0,\lambda}|^2,
\end{align}
which represents the spectrum of eigenenergies excited by the initial state. A peak in the excitation spectrum thus indicates that the initial state more densely excites the continuum eigenstates around that energy (in the sense of a resonance). To obtain a dynamical interpretation of $F_\lambda$, the amplitude of the initial component $\ket{\phi_0}$ as a function of time is given by
\begin{align} \label{eq:generic-decay}
\begin{split}
    &\langle \phi_0\vert e^{-\mathrm{i}\hat{G}\tau}\vert \phi_0\rangle\\
    &\quad{}=\sum_{k}e^{-\mathrm{i}\lambda_k\tau}|c_{0,\lambda_k}|^2+\int \mathrm{d}\lambda \, e^{-\mathrm{i}\lambda\tau} |F_{\lambda}|^2,
    \end{split}
\end{align}
whose second term can be seen as a Fourier transform of the excitation spectrum (which also gives it a transfer-function interpretation). In many cases, when a single discrete state couples to a continuum, $|F_{\lambda}|^2$ takes a Lorenzian-like form, which then implies that the second term in \eqref{eq:generic-decay} represents an exponential ``decay'' out of the initial state (and into the continuum). Taking cues from the physics of intramolecular vibrational energy redistribution, we view this as a ``dissipative'' effect.

When multiple discrete states are coupled to the continuum, Fano interferences can produce asymmetric peaks in $|F_{\lambda}|^2$, i.e., Fano resonances~\cite{Miroshnichenko2010,Limonov2017}, which can have remarkably narrow lineshapes. Under certain conditions, the Fano resonance can become infinitely narrow, indicating that the initial state is exciting an eigenstate with some discrete eigenenergy $\lambda^*$, i.e., a bound state. If $\lambda^*$ lies within the band of continuum eigenstates ($\lambda^*>\lambda_\text{c}$), then that state is an example of a bound state in the continuum (BIC). BICs have have drawn much attention both from scientific and engineering points of view~\cite{Hsu2016} due to their potential for realizing decay-free evolution even in the presence of strong coupling between the initial discrete state to a continuum ``environment'' which would otherwise lead to dissipation.

\section{DISSIPATIVE AND DISPERSIVE LIMITS FOR PUMP POPULATION DYNAMICS} \label{sec:asymptotic}
In this section, we analyze the time evolution of the pump photon population \eqref{eq:evolution} in the limits of dissipative coupling $\xi\rightarrow\infty$ and dispersive coupling $\xi=-\infty$. We show that, in the dissipative limit, the pump population follows an exponential decay, while for the dispersive limit, Rabi-like oscillations with sub-exponential decay occur.

\begin{figure}[h]
    \begin{center}
        \includegraphics[width=0.45\textwidth]{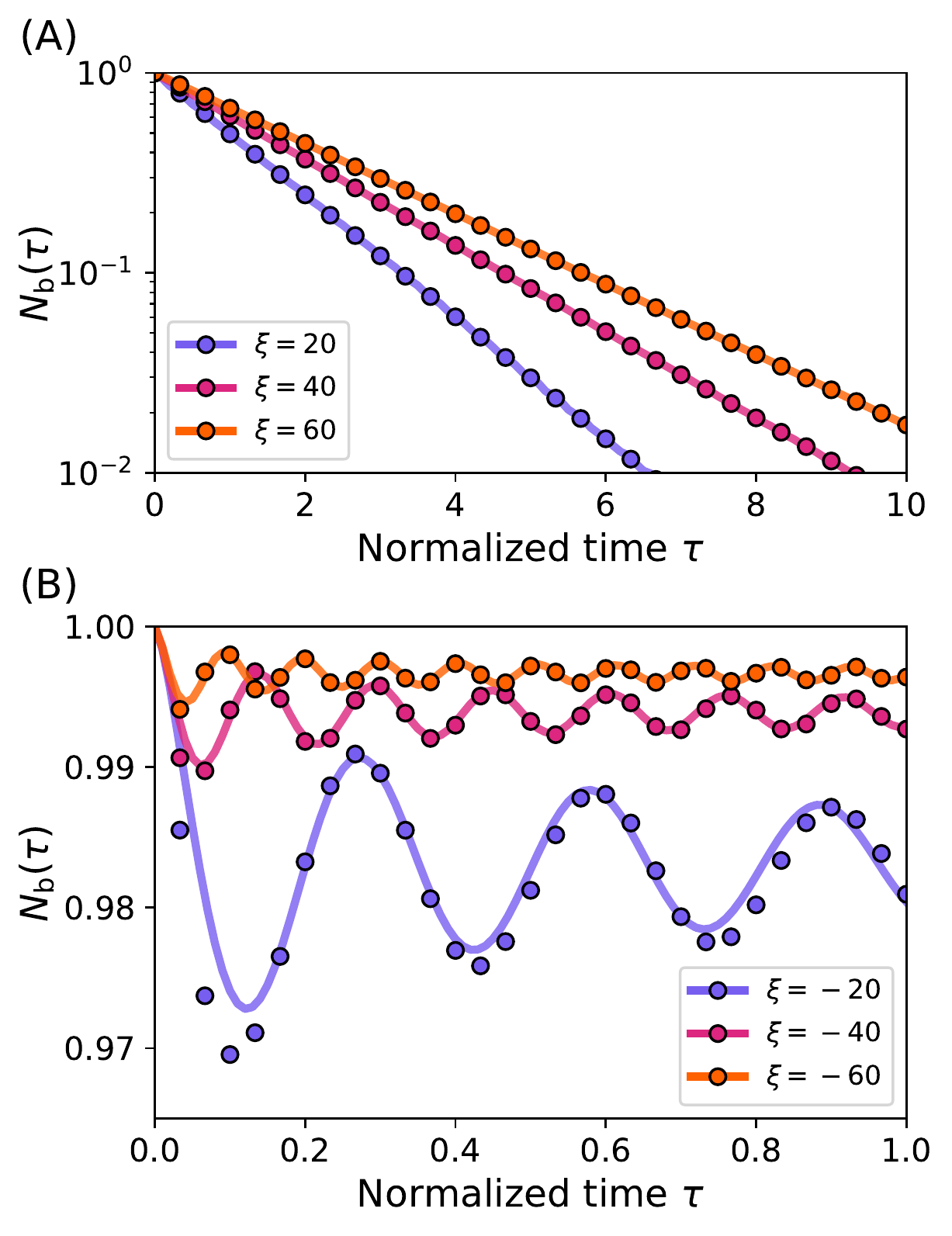}
        \caption{Time evolution of the pump photon number $N_\text{b}(\tau)$ based on the exact numerical expression \eqref{eq:evolution} (solid lines) compared to approximate formulas (circles) valid in the limit of (A) dissipative coupling $\xi\rightarrow\infty$ \eqref{eq:positive_approx}, and (B) dispersive coupling $\xi\rightarrow -\infty$ \eqref{eq:negative_approx}.}
        \label{fig:approximation}
    \end{center}
\end{figure}

First, we consider the dissipative limit $\xi\rightarrow \infty$. In this limit, the optical meson state is mostly composed of signal excitations (see \eqref{eq:cm}), so the pump photon population $N_\text{b}(\tau)$ can be obtained by only considering the contributions from the continuum states. As a result, $N_\text{b}(\tau)$ can be approximated using \eqref{eq:ctau} with $c_\text{M} \ll 1$, which gives
\begin{align}
    \label{eq:cont_integral}
    N_\text{b}(\tau)\approx\left\vert\int^\infty_0 \mathrm{d}\lambda \, c_\lambda^2e^{-\mathrm{i}\lambda\tau}\right\vert^2.
\end{align}
When $\xi\gg1$ holds true, $c_\lambda^2$ can be approximated as a Lorentzian function with full width at half maximum $\pi/\sqrt{\xi}$ centered around $\lambda\approx\xi\gg1$ (see \eqref{eq:clambdacont}). In this case, \eqref{eq:cont_integral} can be seen as a Fourier transform, resulting in a characteristic exponential decay
\begin{align}
    \label{eq:positive_approx}
    N_\text{b}(\tau)\approx \exp(-\tau/\tau_\mathrm{d}),
\end{align}
where $\tau_\text{d}=\sqrt{\xi}/\pi$ is the characteristic time scale for the exponential decay of the pump population. As we show in Fig.~\ref{fig:approximation}, \eqref{eq:positive_approx} agrees well with the exact result \eqref{eq:evolution} when $\xi$ is a large positive value.

Next, we consider the dispersive limit $\xi\rightarrow-\infty$. Based on \eqref{eq:implicit}, the energy of the optical meson can be approximately given as $\lambda_\text{M}\approx -\xi$, which means the contribution to the pump population from the optical meson state is $c_{\text{M}}^2\approx 1-\frac{\pi}{4}(-\xi)^{-\frac{3}{2}}$ using \eqref{eq:cm} and \eqref{eq:ctau}. As a result, we can approximate \eqref{eq:evolution} as
\begin{align}\label{eq:int-dispersive}
    N_\text{b}(\tau)\approx \left\vert 1-\frac{\pi}{4(-\xi)^\frac{3}{2}}+\int^\infty_0 \mathrm{d}\lambda\, c_\lambda^2e^{-\mathrm{i}\lambda\tau}e^{\mathrm{i}\xi \tau}\right\vert^2.
\end{align}
In order to capture the qualitative behavior of $c_\lambda^2$, we use
\begin{align}
    c_\lambda^2=\frac{2\sqrt{\lambda}}{4(\lambda-\xi)^2\lambda+\pi^2}\approx h(\lambda)=\frac{2\sqrt{\lambda}}{4\xi^2\lambda+\pi^2},
\end{align}
where $h(\lambda)$ approximates $c_\lambda^2$ well for $\lambda\ll|\xi|$, which is the region where $c_\lambda^2$ has a peak. Using $h(\lambda)$, we can analytically perform the integral, which yields
\begin{align}
    &\int^\infty_0\mathrm{d}\lambda\, h(\lambda)e^{-\mathrm{i}\lambda\tau}=\frac{\sqrt{2\pi}(1-\mathrm{i})}{4\xi^2\sqrt{\tau}}+\frac{\pi^2\exp\left(\frac{\mathrm{i}\pi^2\tau}{4\xi^2}\right)}{4\xi^3}\nonumber\\
    &\times\left[-1+(1+\mathrm{i})C\left(\sqrt{\frac{\pi\tau}{2\xi^2}}\right)+(1-\mathrm{i})S\left(\sqrt{\frac{\pi\tau}{2\xi^2}}\right)\right]\nonumber\\
    &\approx\frac{\sqrt{2\pi}(1-\mathrm{i})}{4\xi^2\sqrt{\tau}},
\end{align}
where we have ignored terms on the order of $\mathcal{O}(\xi^{-3})$ in the last line. Here, $C(x)$ are $S(x)$ are Fresnel integrals
\begin{align}
    &S(x)=\int^x_0\mathrm{d}t\sin(t^2), &C(x)=\int^x_0\mathrm{d}t\cos(t^2).
\end{align}
Note that $c_\lambda^2$ and $h(\lambda)$ have different asymptotic scalings at $\lambda\rightarrow\infty$, which affects the convergence of the integral involving $h(\lambda)$ at $\tau\rightarrow 0$. For finite $\tau$, however, contribution from these tails of $c_\lambda^2$ and $h(\lambda)$ at large $\lambda$ are washed out by the phase rotations $e^{\mathrm{i}\lambda\tau}$ in the integral, so they contribute negligibly to $N_\text{b}(\tau)$ at finite $\tau$.

Thus, by approximating $c_\lambda^2$ with $h(\lambda)$ in the integral of \eqref{eq:int-dispersive}, the pump population for $\xi\rightarrow-\infty$ can be approximated as
\begin{align}
    \label{eq:negative_approx}
    N_\text{b}(\tau)\approx \left\vert 1-\frac{\pi}{4(-\xi)^\frac{3}{2}}+\frac{\sqrt{\pi}}{2\xi^2\sqrt{\tau}}e^{\mathrm{i}\left(\xi \tau-\frac{\pi}{4}\right)}\right\vert^2,
\end{align}
which exhibits sinusoidal oscillations with frequency $\xi$ with sub-exponential decay. As shown in Fig.~\ref{fig:approximation}, \eqref{eq:negative_approx} reproduces the period and the characteristic decay of the oscillations of the exact result \eqref{eq:evolution} when $\xi$ takes a large negative value.

\section{OFF-DIAGONAL HAMILTONIAN TERMS FOR MULTI-PHOTON PDC} \label{sec:multiplicity}
In this section, we derive an explicit expression for the proportionality factors arising in \eqref{eq:offdiag_n} for the matrix elements of the discrete Hamiltonian in the multi-photon case. The off-diagonal elements between $\ket{\mathcal{M}_{2N-2J},\mathcal{L}_J}$ and $\ket{\mathcal{M}'_{2N-2J'},\mathcal{L}'_{J'}}$ (see Sec.~\ref{sec:general} for notation) are due solely to the nonlinear part of the Hamiltonian $\hat{H}/\hbar\kappa$ (see \eqref{eq:fullhamiltonian})
\begin{align}\label{eq:Hnl-multi}
    \frac{\epsilon^\frac{1}{2}}{2}\sum_{m+n=\ell}\left(\hat{a}_{m}^\dagger\hat{a}_{n}^\dagger\hat{b}_{\ell}+\hat{a}_{m}\hat{a}_{n}\hat{b}_{\ell}^\dagger\right),
\end{align}
which is only nonzero when $|J'-J| = 1$. As a result, we can focus on the case $J'+1=J\geq1$ without loss of generality, corresponding to the action of the first term of \eqref{eq:Hnl-multi}. The action of \eqref{eq:Hnl-multi} in this case is to annihilate a photon from one of the labels of $\mathcal L_J$, say $\widetilde \ell$, and create a pair of photons in two of the labels of $\mathcal M_{2N-2J}$, say $\widetilde m$ and $\widetilde n$, such that $\widetilde m + \widetilde n = \widetilde \ell$. When this procedure produces a resulting set of labels that match those of $\ket{\mathcal{M}'_{2N-2J'},\mathcal{L}'_{J'}}$, we have a nonzero matrix element; in multi-set notation, this condition can be written as
\begin{align}
\begin{split}
    \mathcal{L}_{J}&=\mathcal{L}'_{J'}\uplus\{\widetilde{\ell}\},\\ \mathcal{M}'_{2N-2J'}&=\mathcal{M}_{2N-2J}\uplus\{\widetilde{m},\widetilde{n}\},
    \end{split}
\end{align}
where $\uplus$ is the multiset sum. For such a term, the action of the annihilation and creation operators result in an overall factor of
\begin{align}
\begin{cases}
    \frac{1}{2}\sqrt{\mathfrak{m}_{\widetilde{\ell}} (\mathfrak{m}_{\widetilde{m}}+1)(\mathfrak{m}_{\widetilde{m}}+2)} & \widetilde{m}=\widetilde{n} \\
    \sqrt{\mathfrak{m}_{\widetilde{\ell}} (\mathfrak{m}_{\widetilde{m}}+1)(\mathfrak{m}_{\widetilde{n}}+1)} & \text{otherwise}\\
    \end{cases},
\end{align}
where $\mathfrak{m}_{\widetilde{\ell}}$ is the multiplicity of $\widetilde{\ell}$ in $\mathcal{L}_J$, and $\mathfrak{m}_{\widetilde{m}}$ and $\mathfrak{m}_{\widetilde{n}}$ are the multiplicities of $\widetilde{m}$ and $\widetilde{n}$ in $\mathcal{M}_{2N-2J}$, respectively. This factor is precisely the proportionality factor in \eqref{eq:offdiag_n}.

\section{CONTINUUM HAMILTONIAN FOR THREE-PHOTON GENERATION}
\label{sec:threephoton}
In this section, we derive a continuum-limit Hamiltonian (valid for quantization window length $L \rightarrow\infty$) for the three-photon generation (TPG) process. With the exception of some mathematical technicalities due to the higher-dimensional continuum, the procedure proceeds in the same as for the PDC case (see Secs.~\ref{sec:discrete} and \ref{sec:continuum}). We restrict our attention to the subspace spanned by states dynamically accessible from an initial single-photon dc state $\ket{c_0}=\hat{c}^\dagger_0\ket{0}$. Starting with the complete Hamiltonian \eqref{eq:chi3hamiltonian}, the relevant signal states for TPG from $\ket{c_0}$ are
\begin{align}
    \ket{a_{p_1,p_2}}\propto \hat{a}_{p_1}^\dagger\hat{a}_{p_2}^\dagger\hat{a}_{p_3}^\dagger\ket{0}
\end{align}
where $p_3=-p_1-p_2$ and $p_3\leq p_2\leq p_1$. The momentum cutoff $K_\text{max}$ introduced in the main text leads to a momentum cutoff $P=\sqrt{p_1^2+p_2^2+p_3^2}\leq P_\text{max}=\frac{K_\text{max}L}{2\pi}$ applied to these states. The nonzero matrix elements of the Hamiltonian are
\begin{subequations}
\begin{align}
    \braket{c_0|\hat{H}|c_0}/\hbar\kappa&=\xi\\
    \label{eq:mult_chi3}
    \braket{a_{p_1,p_2}|\hat{H}|a_{p_1,p_2}}/\hbar\kappa&\propto \epsilon^2(p_1^2+p_2^2+p_1p_2)\\
    \braket{a_{p_1,p_2}|\hat{H}|c_0}/\hbar\kappa&=\epsilon
\end{align}
\end{subequations}
where $\kappa=2\chi^2d_\text{a}^{-1}$ is the effective nonlinear coupling rate (independent of $L$), $\epsilon=d_\text{a}/\chi\propto L^{-1}$ is a dimensionless parameter characterizing the quantization length, and $\xi=\delta/\kappa$ is the normalized detuning arising from phase mismatch. The proportionality factor in \eqref{eq:mult_chi3} is
\begin{align}
    \left\{\begin{array}{ll}
        1&\text{for}~p_3<p_2<p_1\\
        \sqrt{6}&\text{for}~p_3=p_2=p_1\\
        \sqrt{2}&\text{otherwise}
    \end{array}\right..
\end{align} 

We introduce two-level lowering operators
\begin{align}
    \hat{w}=\ket{0}\bra{c_0}\quad\text{and}\quad \hat{u}_{p_1,p_2}=\ket{0}\bra{a_{p_1,p_2}},
\end{align}
for annihilating the pump and signal-triplet excitations. Using these operators, we can define an effective hopping Hamiltonian for this subspace, normalized by $\hbar\kappa$, as
\begin{align}
\label{eq:gchidisc}
    \hat{G}&=\xi\hat{w}^\dagger\hat{w}+\!\!\!\sum_{\substack{-p_1-p_2\leq p_2\leq p_1\\P\leq P_\text{max}}} \!\!\!\epsilon^2(p_1^2+p_2^2+p_1p_2)\hat{u}^\dagger_{p_1,p_2}\hat{u}_{p_1,p_2}\nonumber\\
    &\qquad{}+\epsilon(\hat{w}\hat{u}_{p_1,p_2}^\dagger+\hat{w}^\dagger\hat{u}_{p_1,p_2}).
\end{align}

To take a continuum limit, we define $s_i=\epsilon p_i~(i=1,2)$, upon which the operators
\begin{align}
\hat{\phi}_{s_1,s_2}=\epsilon^{-1} \hat{u}_{p_1,p_2}
\end{align}
have a commutation relationship which limit to $\bigl[\hat{\phi}_{s_1,s_2},\hat{\phi}^\dagger_{s_1',s_2'}\bigr]=\delta(s_1-s_1')\delta(s_2-s_2')$ for $L \rightarrow\infty$. Rewriting \eqref{eq:gchidisc} in this limit gives
\begin{align}
\label{eq:gcontchi}
    \hat{G}^\text{cont}&=\xi\hat{w}^\dagger\hat{w}+\int^\infty_0\text{d}s_1\int^{s_1}_{-s_1/2}\text{d}s_2\,\Theta(r_\text{max}-r)\nonumber\\
    &{}\times\left(r^2\hat{\phi}_{s_1,s_2}^\dagger\hat{\phi}_{s_1,s_2}+\hat{w}\hat{\phi}_{s_1,s_2}^\dagger+\hat{w}^\dagger\hat{\phi}_{s_1,s_2}\right),\
\end{align}
where $r^2=s_1^2+s_2^2+s_1s_2$, and $\Theta(x)=(x+|x|)/2x$ is the Heaviside step function. The momentum cutoff in this parametrization takes the form $r_\text{max}=K_\text{max}\zeta/2\sqrt{2}\pi$, where $\zeta=\epsilon L$ (independent of $L$) is the characteristic correlation length.

To take into account the momentum cutoff naturally, we introduce polar coordinates
\begin{align}
&r\cos\theta =\frac{\sqrt{3}}{2}s_1, &r\sin\theta=s_2+\frac{s_1}{2},
\end{align}
where $0\leq r\leq r_\text{max}$ and $0\leq \theta \leq \frac{\pi}{3}$. We can also reparametrize the field annhilation operators as
\begin{align}
\hat{\phi}_{r,\theta}=\sqrt{r}\hat{\phi}_{s_1,s_2}
\end{align}
which obey $\big[\hat{\phi}_{r,\theta},\hat{\phi}^\dagger_{r',\theta'}\big]=\delta(r-r')\delta(\theta-\theta')$. Using these operators, \eqref{eq:gcontchi} can be rewritten as
\begin{align}
\label{eq:gcontchipolar}
    \hat{G}^\text{cont}&=\xi\hat{w}^\dagger\hat{w}+\frac{2}{\sqrt{3}}\int^{r_\text{max}}_0\mathrm{d}r\int^{\pi/3}_0\mathrm{d}\theta\nonumber\\
    &\times\left\{r^2\hat{\phi}^\dagger_{r,\theta}\hat{\phi}_{r,\theta}+\sqrt{r}\left(\hat{\phi}_{r,\theta}\hat{w}^\dagger+\hat{\phi}_{r,\theta}^\dagger\hat{w}\right)\right\}.
\end{align}
The Hamiltonian \eqref{eq:gcontchipolar} has is symmetric in $\theta$, and this symmetry is preserved by any dynamics starting from the initial state $\ket{c_0} = \hat{w}^\dagger\ket{0}$. By discarding the interactions in 
$\hat{G}^\text{cont}$ that do not preserve this symmetry, we can further simplify the continuum Hamiltonian to
\begin{align}
\label{eq:gcontchiS}
&\hat{G}^\text{cont}=\xi\hat{w}^\dagger\hat{w}\\
&+\frac{2}{\sqrt{3}}\int^{r_\text{max}}_0\!\!\mathrm{d}r\left[r^2\hat{\phi}^\dagger_{r}\hat{\phi}_{r}+\sqrt{\frac{\pi r}{3}}\left(\hat{\phi}_{r}\hat{w}^\dagger+\hat{\phi}_{r}^\dagger\hat{w}\right)\right]\nonumber,
\end{align}
where we have defined the ($\theta$-symmetric) signal continuum photon-triplet annhilation operators
\begin{align}
\label{eq:superposition}
\hat{\phi}_r=\sqrt{\frac{3}{\pi}}\int^{\pi/3}_0\mathrm{d}\theta \, \hat{\phi}_{r,\theta}
\end{align}
which obey $\bigl[\hat{\phi}_r,\hat{\phi}^\dagger _{r'}\bigr]=\delta(r-r')$.

\end{appendix}
\bibliography{myfile}
\end{document}